\documentclass[a4paper,11pt]{article}
\usepackage[english]{babel}
\usepackage{t1enc}
\usepackage[utf8]{inputenc}
\usepackage{lmodern}
\usepackage[pdftex]{graphicx}
\usepackage{epstopdf}
\usepackage{amsmath,amssymb}
\usepackage{amsthm}
\usepackage{icomma}
\usepackage{hhline}
\usepackage{array}
\usepackage{cite}
\usepackage[unicode,colorlinks]{hyperref}

\usepackage{booktabs}

\usepackage{mathrsfs}

\usepackage{caption}
\usepackage{subcaption}

\usepackage{feynmp}
\newtheorem{theorem}{Theorem}[section]

\usepackage{pdfpages}

\usepackage{pgfplots}
\pgfplotsset{compat=1.17}

\usepackage[margin=1.0in]{geometry}

\numberwithin{equation}{section}
\numberwithin{table}{section}
\numberwithin{figure}{section}

\usepackage{tikz}
\usetikzlibrary{positioning}
\usetikzlibrary{calc}

\begin{document}

\title{Form factor bootstrap in the thermally perturbed tricritical Ising model}

\author{B. Fitos$^{1,2}$ and G. Tak\'acs$^{1,2,3}$ \\
$^{1}${\small{}{}{}
Department of Theoretical Physics, Institute of
Physics,}\\
 {\small{}{}{}Budapest University of Technology and
Economics,}\\
 {\small{}{}{} M{\H u}egyetem rkp. 3., H-1111 Budapest,
Hungary}\\
 $^{2}${\small{}{}{}
BME-MTA Statistical Field Theory ’Lend\"ulet’ Research
Group,}\\
 {\small{}{}{}Budapest University of Technology and
Economics,}\\
 {\small{}{}{} M{\H u}egyetem rkp. 3., H-1111 Budapest,
Hungary}\\
  $^{3}${\small{}{}{}
MTA-BME Quantum Dynamics and Correlations
Research Group,}\\
 {\small{}{}{}Budapest University of Technology and
Economics,}\\
 {\small{}{}{} M{\H u}egyetem rkp. 3., H-1111 Budapest,
Hungary} }

\date{June 4, 2024}
\maketitle

\begin{abstract}
We derive a systematic construction for form factors of relevant fields in the thermal perturbation of the tricritical Ising model, an integrable model with scattering amplitudes described by the $E_7$ bootstrap. We find a new type of recursive structure encoding the information in the bound state fusion structure, which fully determines the form factors of the perturbing field and the order/disorder fields, for which we present a mathematical proof. Knowledge of these form factors enables the systematic computation of correlation functions and dynamical structure factors in systems whose dynamics is governed by the vicinity of a fixed point in the tricritical Ising universality class.
\end{abstract}

\tableofcontents

\section{Introduction}

The statistical mechanics of integrable low-dimensional magnets has recently attracted renewed interest due to the increased availability of experimental realisations. A well-known example is the material $\text{Co}(\text{NbO}_3)_2$, for which the excitation spectrum obtained by inelastic neutron scattering revealed spectral lines in close correspondence for the golden ratio \cite{2010Sci...327..177C}. This was argued to correspond to a famous exactly solved system investigated by Zamolodchikov \cite{1989IJMPA...4.4235Z}, where the scattering of quasi-particle excitations is described by a factorised $S$-matrix related to the exceptional Lie algebra $E_8$ corresponding to the magnetic deformation of the critical Ising model. Recent THz spectroscopy studies gave a much more detailed map of the excitation spectrum and indicated additional excitation peaks beyond those predicted by the $E_8$ integrable model \cite{2020PhRvB.102j4431A,Amelin:2022ehz}, prompting a proposal to describe the dynamics of the system using a scattering theory related to the $D_8^{(1)}$ affine Lie-algebra, which corresponds to a reflectionless point in the attractive regime of the quantum sine-Gordon model \cite{2024arXiv240211229G,2024arXiv240310785X}. More recently, the full $E_8$ spectrum was observed in the antiferromagnetic spin chain material $\text{BaCo}_2\text{V}_2\text{O}_8$ \cite{Amelin:2022ehz,2020arXiv200513302Z,2021PhRvB.103w5117W,2023arXiv230800249W}.

In the case of both inelastic neutron scattering and THz spectroscopy, the experimental signatures correspond to the measurement of dynamical structure factors, which can be constructed as the Fourier transform of magnetisation two-point functions (for a recent review the reader is referred to \cite{2023JPhA...56L3001L}). For systems described by an integrable quantum field theory, such correlations can be constructed using the form factor bootstrap program \cite{Smirnov}. For the case of the $E_8$ model, these form factors were constructed in \cite{Delfino:1996jr,DMIMMF,2021PhRvB.103w5117W}.

Here, we consider another paradigmatic integrable field theory, the $E_7$ model, which corresponds to the thermal deformation of the tricritical Ising model \cite{MC,FZ}. The form factors for the trace of the energy-momentum tensor were considered in \cite{AMV}; however, the problem of the order parameters has long presented a challenge, with a breakthrough only achieved recently, leading to the computation of dynamical structure factors \cite{2022ScPP...12..162C}. More recently, the form factor bootstrap of the $E_6$ model corresponding to the thermal deformation of the tricritical three-state Potts model was considered in \cite{2024JSMTE2024c3103M} up to two-particle form factors, with the results used to compute universal ratios of the renormalisation group.

The existing results for the $E_7$ stop at the level of two-particle form factors, and a systematic construction is still lacking. In this paper, we present a solution for the form factors of order/disorder fields in the $E_7$ model that works systematically for any number of particles. The method also allows the construction of the higher form factors of the trace of the energy-momentum tensor. This is performed by setting up a recursion scheme to construct all form factors containing only the lightest particle $A_1$, from which all higher ones can be obtained by bootstrap fusion. 

The problem considered here is especially interesting in the light of the considerable recent progress towards an experimental realisation of Ising tricriticality \cite{2019PhRvL.123i0401B,2021PhRvB.104w5109S,2022PhRvB.106k5122S,2023PhRvB.108w5414R,PhysRevLett.132.226502}. Furthermore, our work is also motivated by many recently studied interesting dynamical scenarios, such as kink confinement \cite{2022PhLB..82837008L} and a wide variety of false vacuum decay processes \cite{2022PhRvD.106j5003L}, which result from adding further perturbing fields to the $E_7$ model.

The outline of the paper is as follows. Section \ref{sec:e7} collects the necessary facts on the exact $S$-matrix and form factor equations of the $E_7$ model. We then discuss the solution of all form factors containing only the second lightest particle $A_2$ in Section \ref{sec:A2}, which is a useful stepping stone to the systematic construction of the form factors containing only $A_1$ presented in Section \ref{sec:A1}. In Section \ref{sec:crosschecks}, crosscheck our results with those in \cite{2022ScPP...12..162C} and also give an efficient method for the construction of general form factors, while in Section \ref{sec:discussion}, we present our conclusions. The paper also contains two appendices, with Appendix \ref{sec:building_blocks} giving the building blocks for the form factor Ansatz, while the proof of the completeness of our recursion scheme is given in Appendix \ref{sec:proof}.

\section{The thermally perturbed tricritical Ising model}\label{sec:e7}
\begin{table}
 \bgroup
 \setlength{\tabcolsep}{0.5em}
 \renewcommand{\arraystretch}{1.3}
 \begin{center}
 \begin{tabular}{|c | c | c | c|}
     \hline
       field  & $(\Delta,\Bar{\Delta})$ &parity & quantity  \\
     \hline
     $\mathbb{I}$ & $\displaystyle\left(0,0\right)$ & even & identity \\
     $\sigma$  & $\displaystyle\left(\tfrac{3}{80},\tfrac{3}{80}\right)$  & odd & magnetisation  \\
     $\epsilon$ & $\displaystyle\left(\tfrac{1}{10},\tfrac{1}{10}\right)$  & even &  energy  \\
     $\sigma'$ & $\displaystyle\left(\tfrac{7}{16},\tfrac{7}{16}\right)$  & odd & subleading magnetisation  \\
     $t$ & $\displaystyle\left(\tfrac{3}{5},\tfrac{3}{5}\right)$  &  even & chemical potential  \\
     $\epsilon'$ & $\displaystyle\left(\tfrac{3}{2},\tfrac{3}{2}\right)$ & even & (irrelevant)  \\
    \hline
  \end{tabular} 
  \end{center}
  \egroup
  \caption{Primary fields of the TIM with their conformal weights $(\Delta,\Bar{\Delta})$, parity under $\mathbb{Z}_2$ and the corresponding physical quantity.} 
  \label{tab:TIM}
\end{table} 
The $E_7$ quantum field theory is the thermal perturbation of the tricritical Ising model, a minimal conformal model with central charge $c=7/10$. The local primary fields of the model are shown in Table \ref{tab:TIM}. The Euclidean version of the above CFT arises at the tricritical point of the Blume--Capel model \cite{BC1,BC2} with the Hamiltonian
\begin{equation}
    H_\text{BCM}(\{s_i,t_i\})=-J\sum_{\langle i,j\rangle}s_i s_j t_i t_j+\Omega\sum_i t_i+K \sum_{\langle i,j\rangle} t_i t_j
    -H\sum_i s_i t_i-H'\sum_{\langle i,j\rangle} \left(s_i t_i t_j+s_j t_j t_i\right)
\label{eq:BC}
\end{equation}
which is a version of the two-dimensional Ising magnet composed of spins $s_i=\pm 1$ with vacancies described by the variables $t_i=0,1$. The indices $i,j$ run over the two-dimensional lattice with $\langle i,j\rangle$ denoting nearest neighbours. The parameter $\Omega$ is the chemical potential for the vacancies, $H$ and $H'$ are two external magnetic fields coupled to two relevant order parameter fields, while $K$ is an (irrelevant) nearest neighbour interaction between the vacancies. The real-time (Minkowski) version of the CFT describes the dynamics of the quantum tricritical Ising spin chain with Hamiltonian \cite{vonGehlen:1989yn}
\begin{equation}
    \hat{H}_\text{TIC}=
    -\mathcal{J}\sum_{i} \left\{S_i^z S_{i+1}^z - \alpha (S_i^z)^2 -\beta S_i^x - \gamma  (S_i^x)^2 -h S_i^z \right \}
    \label{eq:TIMC}
\end{equation}
at its tricritical point, where the spin-$1$ degrees of freedom are 
\begin{equation}
    S^z=
    \begin{pmatrix}
    1 & 0 & 0 \\ 0 & 0 & 0 \\ 0 & 0 & -1
    \end{pmatrix}\quad
    S^x=\frac{1}{\sqrt{2}}
    \begin{pmatrix}
    0 & 1 & 0 \\ 1 & 0 & 1 \\ 0 & 1 & 0
    \end{pmatrix}\,.
\end{equation}
Both models \eqref{eq:BC} and \eqref{eq:TIMC} have a $\mathbb{Z}_2$ symmetry, which can be broken spontaneously, signalled by a nonzero expectation value of magnetisation in the ferromagnetic phase, while the expectation value vanishes in the paramagnetic one. The two phases are separated by a line, part of which corresponds to a first-order, while the other part to a second-order phase transitions, with a tricritical point separating the two. The two phases are related by a Kramers--Wannier duality transformation, which corresponds to the following transformation of the even fields:
\begin{equation}
    \epsilon\rightarrow -\epsilon\quad ,\quad
    \epsilon'\rightarrow -\epsilon'\quad ,\quad
    t\rightarrow t\,,
\label{eq:KW}\end{equation}
with the sign change of the energy field corresponds to interchanging the two phases. Under duality, the order parameter field $\sigma$ and $\sigma'$ transform into the disorder fields $\mu$ and $\mu'$ of the same conformal dimensions, which are, however, non-local with respect to $\sigma$ and $\sigma'$. Due to the Kramers--Wannier duality, it is enough to determine the matrix elements of the local operators in one of the phases since their values in the other phase can be obtained using the transformations \eqref{eq:KW}.  We refer the interested reader to \cite{2023JPhA...56L3001L,2022ScPP...12..162C} for more details. 

The thermal perturbation of the tricritical Ising CFT is given by the formal action
\begin{align}
    \mathcal{A} = \mathcal{A}_\text{TIM} + g\int d^2x \, \epsilon(x)
\end{align}
with the positive (negative) value of the coupling $g$ corresponding to the paramagnetic (ferromagnetic) phases. The resulting theory has a mass gap with seven massive excitations listed in Table \ref{tab:e7_particles}. In the ferromagnetic phase, the odd excitations are topologically charged kink states and their neutral bound states, while in the paramagnetic phase, all excitations are topologically trivial. 

The off-critical theory is integrable, with the spins of the conserved charges related to Coxeter exponents of the exceptional algebra ${E}_7$, which also governs the two-particle scattering matrix \cite{MC,FZ}. Due to integrability, all scattering processes factorise into the product of independent two-particle scattering processes \cite{1979AnPhy.120..253Z}. The full spectrum consists of seven massive excitations $A_n$ with masses $m_n$ ($n=1,\dots,7$), with the mass gap given by \cite{FATEEV199445}
\begin{align}
    \begin{aligned} \label{Eq:massgap}
        m_1 &= \frac{2\Gamma\left(\frac{2}{9}\right)}{\Gamma\left(\frac{2}{3}\right)\Gamma\left(\frac{5}{9}\right)}\left(\frac{4\pi^2\Gamma\left(\frac{2}{5}\right)\Gamma^3\left(\frac{4}{5}\right)}{\Gamma^3\left(\frac{1}{5}\right)\Gamma\left(\frac{3}{5}\right)}\right)^{5/18}|g|^{5/9}\\
        &= 3.745372836243954\ldots\,|g|^{5/9}\,.
    \end{aligned}
\end{align}
The mass of all other excitations is determined by the $S$-matrix bootstrap. The two-particle amplitudes involving $A_a$ and $A_b$ are denoted by $S_{ab}$ and are summarised in Appendix \ref{sec:building_blocks}.
\begin{table} 
    \bgroup
    \setlength{\tabcolsep}{0.5em}
  \begin{center}
  \begin{tabular}{|l|l|}
  \hline
  mass & parity\\
  \hline
   $m_1$ & \textcolor{red}{odd} \\
   $m_2=2m_1\cos(5\pi/18)$ & even \\
   $m_3=2m_1\cos(\pi/9)$ & \textcolor{red}{odd} \\
   $m_4=2m_1\cos(\pi/18)$ & even \\
   $m_5=4m_1\cos(\pi/18)\cos(5\pi/18)$ & even \\
   $m_6=4m_1\cos(2\pi/9)\cos(\pi/9)$ & \textcolor{red}{odd} \\
   $m_7=4m_1\cos(\pi/18)\cos(\pi/9)$ & even\\
  \hline
  \end{tabular}
  \end{center}
    \egroup
 \caption{Spectrum of the thermal deformation of the tricritical Ising Model.} \label{tab:e7_particles} 
\end{table}
The trace of the energy-momentum tensor can be expressed in terms of the perturbing field
\begin{align}
    \begin{aligned}
        \Theta(x) &= 4\pi (1 - \Delta_\epsilon) g \epsilon(x)= \frac{18\pi}{5} g \epsilon(x)\,.
    \end{aligned}
\end{align}
Therefore, determining the matrix elements of the $\Theta$ field also gives the matrix elements of the $\epsilon$ field.

\section{Form factor bootstrap}\label{sec:ffe7}

Here we briefly summarise the form factor bootstrap equations used in the sequel. Since in the $E_7$ model, all particles are non-degenerate in their mass, we only give the equations for this particular case, which results in substantial simplifications compared to the general case. Consequently, the scattering theory is diagonal, i.e., all two-particle $S$-matrices are pure phase shifts. It also implies that all particles are identical to their anti-particles (self-conjugate).

\subsection{Form factor equations}

The form factors of a local field $\phi(x)$ are defined as the matrix element between the vacuum and a general multi-particle state
\begin{align}
    F^\phi_{a_1,\ldots,a_n}(\vartheta_1, \ldots, \vartheta_n) &= \langle 0|\phi(0)|A_{a_1}(\vartheta_1),\ldots,A_{a_n}(\vartheta_n)\rangle\,.
\end{align}
The multi-particle states
\begin{equation}
    |A_{a_1}(\vartheta_1),\ldots,A_{a_n}(\vartheta_n)\rangle
\end{equation}
contain particles of species $a_k$ and rapidity $\vartheta_k$, where the latter specifies the energies and momenta of the particles according to the relations $E_k=m_{a_k}\cosh\vartheta_k$ and $P_k=m_{a_k}\sinh\vartheta_k$.

For a relativistic integrable QFT, the form factors continued to complex values of the rapidities are analytic functions apart from isolated poles and satisfy a system of functional equations (for a review c.f.~\cite{Smirnov}) listed below: 
\paragraph{Lorentz invariance:}
\begin{align}
\label{eq:FF_Lorentz}
    \begin{aligned}
        F^\phi_{a_1,\ldots,a_n}(\vartheta_1 + \lambda, \ldots, \vartheta_n + \lambda) &= e^{s_\phi\lambda}F^\phi_{a_1,\ldots,a_n}(\vartheta_1, \ldots, \vartheta_n)\,,
    \end{aligned}
\end{align}
where $s_\phi=\Delta_\phi-\Bar{\Delta}_\phi$ is the Lorentz spin of the local field $\phi$. For scalar fields $s_\phi=0$, and Lorentz invariance implies that the form factors only depend on the rapidity differences.
\paragraph{Monodromy properties:}
\begin{align}
    F^\phi_{...,a_i,a_{i+1},...}(...,\vartheta_i, \vartheta_{i+1},...) &= S_{a_ia_{i+1}}(\vartheta_i-\vartheta_{i+1})F^\phi_{...,a_{i+1},a_i,...}(...,\vartheta_{i+1},\vartheta_i,...)\,,\label{eqn:FF_exchange}\\
    F^\phi_{a_1, a_2,...,a_n}(\vartheta_1 + 2\pi i, ...,\vartheta_n) &= e^{2\pi i\omega_a}F^\phi_{a_2,...,a_n,a_1}(\vartheta_2,...,\vartheta_n,\vartheta_1)\,.\label{eqn:FF_periodic}
\end{align}
where $\omega_a$ is the mutual semi-locality index of the operator $\phi$ with respect to the particle $a$.
\paragraph{Kinematic poles:}
\begin{align}
    \begin{aligned} \label{Eq:KPE}
    -i\lim_{\Tilde{\vartheta}\rightarrow\vartheta} (\Tilde{\vartheta} - \vartheta) &F^\phi_{a,\Bar{a},a_1,...,a_n}(\Tilde{\vartheta}+i\pi, \vartheta, \vartheta_1,...,\vartheta_n)\\
    &= \left(1 -e^{2\pi i\omega_a}\prod_{k=1}^n S_{aa_k}(\vartheta-\vartheta_j)\right)F^\phi_{a_1,...,a_n}(\vartheta_1,...,\vartheta_n)\,.
    \end{aligned}
\end{align}
\paragraph{Bound state poles:}
Whenever particle $c$ occurs as a bound state in the $ab$ scattering amplitude with the pole contribution 
\begin{align} \label{eq:bstatepole}
    S_{ab}(\vartheta_a-\vartheta_b)= \frac{i\left|\Gamma_{ab}^c\right|^2}{\vartheta_a-\vartheta_b-i u_{ab}^c}+\text{reg. terms}\,,
\end{align}
the form factors have corresponding poles
\begin{align} \label{Eq:BSP}
    -i\lim_{\vartheta_{ab}\rightarrow iu_{ab}^c}(\vartheta_{ab} - iu_{ab}^c)F^\phi_{a,b,a_1,...,a_n}(\vartheta_a,\vartheta_b,\vartheta_1,...,\vartheta_n) = \Gamma_{ab}^cF^\phi_{c,a_1,...,a_n}(\vartheta_c, \vartheta_1,...,\vartheta_n)\,,
\end{align}
where $\vartheta_c=\vartheta_a-i\Bar{u}_{ac}^b=\vartheta_b+i\Bar{u}_{bc}^a$, and $\Bar{u}$ denotes the supplementary angle $\Bar{u}= \pi - u$.

\subsection{Additional properties}

Two additional properties are satisfied by form factors of scaling fields, which can be used for operator identification.

\paragraph{Bound on the asymptotic growth:}
Form factors of a scalar scaling operator $\phi(x)$ with conformal weight $\Delta_\phi$ satisfy the following bound for large rapidities \cite{DMIMMF}:
\begin{align}\label{Eq:asymgrowth}
    \lim_{|\vartheta_i|\rightarrow\infty}F^\phi_{a_1,...,a_n}(\vartheta_1,...,\vartheta_n) \sim e^{y_\phi|\vartheta_i|}
\end{align}
where $y_\phi\leq \Delta_\phi$.

\paragraph{Cluster property:}
Form factors of relevant scaling fields also satisfy the following asymptotic factorisation \cite{Koubek:1993ke,AMVcluster,DSC,1997NuPhB.497..589A}
\begin{align}
\label{eq:clusterFF}
    \begin{aligned}
    \lim_{\alpha\rightarrow\infty} &F_{r+l}^{\phi_a}(\vartheta_1 + \alpha,\ldots,\vartheta_r+\alpha, \vartheta_{r+1}, \ldots,\vartheta_{r+l})=\\
    &\text{(phase factor)} F^{\phi_b}_r(\vartheta_1,\ldots,\vartheta_r)F^{\phi_c}_l(\vartheta_{r+1},\ldots\vartheta_{r+l})\,,
    \end{aligned}
\end{align}
where $\phi_a, \phi_b, \phi_c$ are relevant fields of the same conformal dimension normalised to have expectation value $1$, the phase factor depends on the relative phases of the multi-particle states. For a given $\phi_a$, the operators $\phi_b$ and $\phi_c$ occurring in \eqref{eq:clusterFF} depend on the symmetry properties of the fields, as we discuss later.

\subsection{Form factor Ansatz}

The general solution of the form factor equations (\ref{eq:FF_Lorentz},\ref{eqn:FF_exchange},\ref{eqn:FF_periodic}) can be written in the form
\begin{align} \label{Eq:genAnsatz}
    F_{a_1,\ldots,a_n}^\phi(\vartheta_1,\ldots,\vartheta_n) &= \Tilde{Q}^\phi_{a_1,\ldots,a_n}(\vartheta_1,\ldots,\vartheta_n)\prod_{k<l}^n\frac{F_{a_ka_l}^\text{min}(\vartheta_k-\vartheta_l)}{D_{a_ka_l}(\vartheta_k-\vartheta_l)(e^{\vartheta_k}+e^{\vartheta_l})^{\delta_{a_ka_l}}}\,,
\end{align}
which contains the following ingredients:
\begin{itemize}
    \item The minimal two-particle form factors $F_{ab}^\text{min}(\vartheta)$ which solve the two-particle monodromy equations and are free of poles and zeros in the strip $\mathcal{S} = \{ \text{Im }\vartheta \in (0,\pi) \}$:
    \begin{align} \label{Eq:minFFSrel}
        F^\phi_{ab}(\vartheta) &= S_{ab}(\vartheta)F^\phi_{ab}(-\vartheta)\,,\\
        F^\phi_{ab}(i\pi+\vartheta) &= F^\phi_{ab}(i\pi-\vartheta)\,.
    \end{align}
    Their solution is unique up to normalisation \cite{PhysRevD.19.2477}; for $E_7$ they are given in Appendix \ref{sec:building_blocks}.
    \item The bound state pole factors $D_{ab}(\vartheta)$ which encode the singularity structure resulting from the bound state singularity equation \eqref{Eq:BSP}, given explicitly in Appendix \ref{sec:building_blocks}. Additionally, the denominator factors 
    \begin{equation}
        (e^{\vartheta_k}+e^{\vartheta_l})^{\delta_{a_ka_l}}
    \end{equation}
    encode the kinematic singularity structure according to \eqref{Eq:KPE}.
    \item The holomorphic function $\Tilde{Q}^\phi_{a_1,\ldots,a_n}(\vartheta_1,\ldots,\vartheta_n)$ which encodes the information about the specific operator $\phi$, and is fully symmetric under permutations of the particles. The Lorentz invariance \eqref{eq:FF_Lorentz} implies the property
    \begin{equation}
        \Tilde{Q}^\phi_{a_1,\ldots,a_n}(\vartheta_1+\lambda,\ldots,\vartheta_n+\lambda)=
        \exp\left\{
        \left(s_\phi+\sum_{k<l}^n\delta_{a_ka_l}\right)\lambda
        \right\}
        \Tilde{Q}^\phi_{a_1,\ldots,a_n}(\vartheta_1,\ldots,\vartheta_n)\,,
    \label{eq:Qtilde_Lorentz}
    \end{equation}
    while the monodromy equation \eqref{eqn:FF_periodic} implies
    \begin{equation}
        \Tilde{Q}^\phi_{a_1,\ldots,a_n}(\vartheta_1,\ldots,\vartheta_n+2\pi i)=
        e^{2\pi i\omega_{a_n}}\Tilde{Q}^\phi_{a_1,\ldots,a_n}(\vartheta_1,\ldots,\vartheta_n)\,.
    \label{eq:Qtilde_monodromy}
    \end{equation}
\end{itemize}
Additionally, matching the residues as required by the kinematic and bound state singularity equations \eqref{Eq:KPE} and \eqref{Eq:BSP} implies recursion relations for the functions $\Tilde{Q}$, and the task of constructing the form factors is therefore reduced to the solving the resulting system.

To find the solution for a particular operator, it is essential to note that the bound state singularity equation allows reconstructing all form factors starting from ones containing only the particle $A_1$, since the bootstrap structure implies that every particle as a bound state of two or more copies of $A_1$. Therefore, it is only necessary to construct the form factors containing only $A_1$, which we call the $A_1$ tower. This leaves the task of determining the functions $\Tilde{Q}^\phi_{1,\ldots,1}$, which is discussed later. It also turns out to be helpful to consider the form factors containing only the particle $A_2$, i.e., the $A_2$ tower, and indeed, we consider it before presenting the solution for the $A_1$ tower.

Finally, we recall from Section \ref{sec:e7} that it is sufficient to determine the form factors in the paramagnetic phase since the corresponding form factors in the ferromagnetic phase can be obtained using the Kramers--Wannier duality \eqref{eq:KW}. Therefore, in all our subsequent calculations, we restrict ourselves to the paramagnetic phase, for which the order fields $\sigma$ and $\sigma'$ have vanishing vacuum expectations, in contrast to the disorder fields $\mu$ and $\mu'$ whose vacuum expectation values do not vanish. Their exact values have been determined in Ref.~\cite{russian2} as
\begin{align}
    \langle\mu\rangle &= \pm 1.59427\ldots\, |g|^{1/24}\,,\\
    \langle\mu'\rangle &= \pm 2.45205\ldots\, |g|^{35/72}\,.
\label{eq:exact_muvevs}\end{align}
We remark that the sign of the $\mu$ and $\mu'$ disorder fields are not well-defined, as reflected in the sign ambiguity in the vacuum expectation values. This ambiguity appears in their form factors as well: the disorder fields are semi-local $\mathbb{Z}_2$ twist fields in the paramagnetic phase, and therefore, their form factors pick up a minus sign upon exchange with odd particles such as $A_1$. Under Kramers--Wannier duality, this ambiguity corresponds to the dependence of the sign of the order parameters expectation values on the vacuum state which is doubly degenerate due to the spontaneous breaking of the $\mathbb{Z}_2$ spin symmetry.

We further note that the operators $\mu$ and $\mu'$ are even, while $\sigma$ and $\sigma'$ are odd under the unbroken $\mathbb{Z}_2$ symmetry in the paramagnetic phase, which leads to selection rules for their matrix elements according to the parity of excitations given in Table \ref{tab:e7_particles}.

\section{Solving the $A_2$ tower}\label{sec:A2}

With a state containing only $A_2$ particles, which are even, the form factors of the order fields $\sigma$ and $\sigma'$ vanish. Therefore, here we consider form factors of the trace of the stress-energy tensor $\Theta$ (which is proportional to the perturbing field) and the disorder fields $\mu$ and $\mu'$. The asymptotic growth bound for each field is satisfied with an exponent $y_\phi<1/2$. 

The Ansatz for their $A_2$ tower form factors is based on eq.~\eqref{Eq:genAnsatz}  
\begin{align}
    \begin{aligned} \label{Eq:FF2nAnsatz}
        F^\phi_{\textbf{2}^n}(\vartheta_1, \ldots, \vartheta_{n}) &\equiv F^\phi_{\underbrace{\textbf{2},\ldots, \textbf{2}}_{n}}(\vartheta_1, \ldots, \vartheta_{n})\\
        &= H_{\textbf{2}^n}^\phi\frac{Q^\phi_{\textbf{2}^n}(x_1,\ldots,x_{n})}{(x_1\cdot\ldots\cdot x_{n})^{n-1}}\prod_{k<l}^{n}\frac{F^\text{min}_{22}(\vartheta_k-\vartheta_l)}{D_{22}(\vartheta_k-\vartheta_l)(x_k+x_l)}\,,
    \end{aligned}
\end{align}
where $x_k = e^{\vartheta_k}$, and we rewrote the function $\Tilde Q$ by factoring out a convenient normalisation factor  
\begin{align}
    \begin{aligned}
        H_{\textbf{2}^n}^\phi = F_2^\phi\kappa_{22}^{-(n-1)^2/4}\,,
    \end{aligned}
\end{align}
where, for later convenience, we introduced the notation
\begin{align}
    \begin{aligned}\label{Eq:kappa}
        \kappa_{ab} &= 4^{-\delta_{ab}}\prod_{\alpha\in\mathcal{A}_{ab}}\left(8g_\alpha(0)\frac{\cos^4\frac{\alpha\pi}{2}}{\sin(\alpha\pi)}\right)^{p_{ab}(\alpha)}\,.
    \end{aligned}
\end{align}
We also factored out a denominator $(x_1\cdots\ldots\cdot x_n)^{n-1}$ which was chosen to ensure that $Q^\phi_{\textbf{2}^n}$ is a homogeneous symmetric polynomial in the $x_k$'s. For the $n=2$ case, it is chosen so that the asymptotic growth constrains $Q_{\textbf{2}^1}^\phi(x)$ to be a constant. For form factors with $n>2$, the form of this denominator is fixed by taking it as a power of $x_1\dots x_n$, which ensures the absence of any singularities, and its exponent is fixed by taking the minimal one such that the kinematic and bound state singularity equations (\ref{Eq:KPE},\ref{Eq:BSP}) lead to polynomial recursion relations between the $Q^\phi_{\textbf{2}^n}$. Note that the particle $A_2$ is local with respect to both the stress-energy tensor $\Theta$ and the disorder fields $\mu$ and $\mu'$.

The partial degrees of the polynomials $Q^\phi_{\textbf{2}^n}$ are restricted by Lorentz invariance \eqref{eq:FF_Lorentz}, by the finiteness of the left-hand side of the clustering property \eqref{eq:clusterFF}. The resulting restrictions are
\begin{align}
    \deg_1\left(Q^\phi_{\textbf{2}^n}\right) &\leq 3n - 3\,,\nonumber\\
    &\,\,\,\vdots\nonumber\\
    \deg_k\left(Q^\phi_{\textbf{2}^n}\right) &\leq 3k\left(n-\frac{k}{2}-\frac{1}{2}\right)\,,\nonumber\\
    &\,\,\,\vdots\nonumber\\ 
    \deg\left(Q^\phi_{\textbf{2}^n}\right) &= \frac{3}{2}n(n-1)\,,
    \label{Eq:totDegFFT2}
\end{align}
where $\deg_k$ denotes the $k$-th partial degree (the total degree of the polynomial in its first $k$ arguments), and $\deg$ denotes the total degree fixed by Lorentz invariance.

A generic $Q^\phi_{\textbf{2}^n}$ polynomial can be expressed in the algebraic basis of the ring of symmetric polynomials given by the elementary symmetric polynomials
\begin{align}
    \sigma_k(x_1,\ldots,x_n) &= \sum_{l_1<\ldots<l_k}^n\prod_{j=1}^kx_{l_j}\,.
\end{align}
We also introduce the shorthand notation 
\begin{align}
    \sigma_\lambda = \prod_i \sigma_{\lambda_i}
\end{align}
where $\lambda$ is a multi-index $(\lambda_1, \ldots, \lambda_l)$, where $\lambda_i \geq \lambda_j$ for $i < j$. The polynomial $Q^\phi_{\textbf{2}^n}$ can be written uniquely as linear combination
\begin{align} \label{Eq:Q2nLinComb}
    Q^\phi_{\textbf{2}^n} = \sum_{\lambda} C^\phi_\lambda \sigma_\lambda\,,
\end{align}
where the sum runs over the multi-indices satisfying
\begin{align} \label{Eq:partToDegTotA2n}
    \sum_i \lambda_i = \deg\left(Q^\phi_{\textbf{2}^n}\right)
\end{align}
and
\begin{align} \label{Eq:partToDegkA2n}
    \sum_i \min(\lambda_i, k) \leq \deg_k\left(Q^\phi_{\textbf{2}^n}\right)
\end{align}
for every $k$. The coefficients $C_\lambda$ are restricted by the recursion relations resulting from the kinematic and bound state singularity equations.

\subsection{Recursion relations}

The kinematic singularity equation \eqref{Eq:KPE} gives the following recursion relation for the $Q^\phi_{\textbf{2}^n}$ polynomials
\begin{align}
    \begin{aligned} \label{Eq:FFT2KP}
        Q^\phi_{\textbf{2}^{n+2}}(-x,x,x_1,\ldots,x_n) =ix^3\cdot U_{\textbf{2}|\textbf{2}^n}(x|\{x_k\})Q^\phi_{\textbf{2}^n}(x_1,\ldots,x_n)\,,
    \end{aligned}
\end{align}
where
\begin{align}
    \begin{aligned}
        U_{\textbf{2}|\textbf{2}^n}(x|\{x_k\}) &= \prod_{k=1}^n\prod_{\alpha\in\mathcal{A}_{22}}[\alpha]_k[-\bar{\alpha}]_k -\prod_{k=1}^n\prod_{\alpha\in\mathcal{A}_{22}}[-\alpha]_k[\bar{\alpha}]_k\,,
    \end{aligned}
\end{align}
and we introduced the notation
\begin{align}
    [\alpha]_k = x - e^{i\pi\alpha}x_k\quad\text{and} \quad [\bar{\alpha}]_k = x + e^{i\pi\alpha}x_k\,.
\end{align}
In addition, the self-fusion property $A_2\times A_2 \rightarrow A_2$ leads to the following recursion relation from the bound state singularity equation \eqref{Eq:BSP}:
\begin{align} \label{Eq:FFT2BSP}
    Q^\phi_{\textbf{2}^{n+1}}(\varphi x, \varphi^{-1}x, x_2,\ldots,x_n) &= \Gamma_{22}^2C_{22\rightarrow 2} x^3 \Lambda_{22\rightarrow 2}(x|\{x_k\}) Q^\phi_{\textbf{2}^n}(x,x_2,\ldots,x_n)\,,
\end{align}
where $\varphi = e^{i\pi/3}$,
\begin{align}
    |\Gamma_{22}^2| &= \sqrt{2\sqrt{3}\cot\left(\frac{\pi}{18}\right)\cot^2\left(\frac{\pi}{9}\right)\tan\left(\frac{2\pi}{9}\right)}\,,\nonumber\\
    C_{22\rightarrow 2} &= \left(1+2\cos\frac{\pi}{9}\right)\left(1+2\sin\frac{\pi}{18}\right)\tan\frac{\pi}{18}\,,\nonumber\\
    \Lambda_{22\rightarrow 2}(x|\{x_k\}) &= \prod_{k=2}^n\left(x - e^{-14i\pi/18}x_k\right)\left(x - e^{14i\pi/18}x_k\right)\left(x + x_k\right)\,.
\end{align}

\subsection{The minimal kernel}
We now consider the question of the extent to which eqs.~\eqref{Eq:FFT2KP} and \eqref{Eq:FFT2BSP} determine the coefficients $C^\phi_\lambda$. Assume that $Q^{(1)}_{\textbf{2}^n}$ and $Q^{(2)}_{\textbf{2}^n}$ are two solutions of the recursion relations. Then their difference $\mathcal{K}_n = Q^{(1)}_{\textbf{2}^n} - Q^{(2)}_{\textbf{2}^n}$ is a mutual kernel of the operators
\begin{equation} 
    (\mathcal{R}^\text{KP}_{2}\mathcal{K}_n)(x|x_3,\ldots,x_n) = \mathcal{K}_n(-x,x,x_3,\ldots,x_n)
    \label{Eq:FFT2KerKP}
\end{equation}
and
\begin{equation}
    (\mathcal{R}^\text{BSP}_{2 2\rightarrow 2}\mathcal{K}_n)(x|x_3,\ldots,x_n) = \mathcal{K}_n(\varphi x,\varphi^{-1}x,x_3,\ldots,x_n)
    \label{Eq:FFT2KerBSP}
\end{equation}
representing the left-hand sides of eqs.~\eqref{Eq:FFT2KP} and \eqref{Eq:FFT2BSP}, respectively. There can generally be many such kernels spanning a linear subspace in the space of the $C_\lambda$ coefficients. The dimension of the kernel space is identical to the number of undetermined parameters.

The kernel is a symmetric homogeneous polynomial respecting the partial degree restrictions in eq.~\eqref{Eq:totDegFFT2}. Since eq.~\eqref{Eq:FFT2KerKP} brings $\mathcal{K}_n$ to zero, $\mathcal{K}_n$ has roots whenever $x_k + x_l = 0$ for any $k < l$. Similarly, as $\mathcal{K}_n$ is a kernel of eq.~\eqref{Eq:FFT2KerBSP}, $\mathcal{K}_n$ has roots whenever $x_k = \varphi^{\pm 2}x_l$ for any $k < l$. Therefore every $\mathcal{K}_n$ kernel should include the
\begin{align}
    \mathcal{K}^\text{min}_{n}(x_1,\ldots,x_n) &= \prod_{k<l}^n(x_k+x_l)(x_k-\varphi^2 x_l)(x_l-\varphi^2 x_k)
\end{align}
minimal kernel. The partial degrees of the minimal kernel is
\begin{align}
    \begin{aligned}
        \deg_k\left(\mathcal{K}^\text{min}_{n}\right) = 3k\left(n - \frac{k}{2} - \frac{1}{2}\right)\,.
    \end{aligned}
\end{align}
Comparing degrees to eq.~\eqref{Eq:totDegFFT2}, we find that the only kernel that satisfies the degree conditions is a constant multiple of $\mathcal{K}^\text{min}$, which means that the kernel subspace is one dimensional at every level $n$.

\subsection{Clustering property}

We utilize the clustering property \eqref{eq:clusterFF} for the case $A_2^{n+1} \rightarrow A_2 \times A_2^n$ to fix the remaining degree of freedom. Since $\Theta$, $\mu$ and $\mu'$ are the only even local fields with their given conformal weights, their form factors must cluster among themselves, giving the relation
\begin{align}
    \begin{aligned} \label{Eq:FFT2Clus}
        \lim_{X\rightarrow \infty} Q^\phi_{\textbf{2}^{n+1}}(X, x_1,\ldots, x_n) X^{-3n} &= \frac{F_2^\phi\kappa_{22}^{-1/4}}{F_0^\phi}\cdot Q^\phi_{\textbf{2}^n}(x_1,\ldots x_n)\,,
    \end{aligned}
\end{align}
where $F_0^\phi$ denotes the vacuum expectation value of the field $\phi$. This equation fixes the coefficients corresponding to terms with maximal (first) partial degree (cf.~eq.~\eqref{Eq:totDegFFT2}). Since the minimal kernel is also maximal in the first partial degree, the clustering property fixes the remaining degree of freedom in the recursion. Therefore the $A_2$ form factor tower can be obtained systematically by starting from $Q^\phi_{\textbf{2}^1} \equiv 1$ and recursively applying eqs.~\eqref{Eq:FFT2KP}, \eqref{Eq:FFT2BSP}, and \eqref{Eq:FFT2Clus}.

The case $n = 2$ is straightforward since, according to the partial degree restrictions, the $Q^\phi_{\textbf{2}^2}$ polynomial can only contain two terms
\begin{align} \label{Eq:generalQ22}
    Q^\phi_{\textbf{2}^2}(x_1,x_2) = C^\phi_{\textbf{2}^2,(2,1)}\sigma_2\sigma_1 + C^\phi_{\textbf{2}^2,(1,1,1)}\sigma_1^3\,.
\end{align}
Note that each term contains at least one $\sigma_1$ factor; as a result, the two-particle form factor does not have a kinematic pole. The second term is fixed by the clustering
\begin{align} \label{Eq:C22v111}
    C^\phi_{\textbf{2}^2,(1,1,1)} &= \frac{F_2^\phi\kappa_{22}^{-1/4}}{F_0^\phi}\,,
\end{align}
and the bound state recursion \eqref{Eq:FFT2BSP} fixes the other one
\begin{align} \label{Eq:C22v21}
    C^\phi_{\textbf{2}^2,(2,1)} &= \frac{\Gamma_{22}^2C_{22\rightarrow 2}}{2\cos\frac{6\pi}{18}} - \frac{F_2^\phi\kappa_{22}^{-1/4}}{F_0^\phi}\left(2\cos\frac{6\pi}{18}\right)^2\,.
\end{align}
In practice, it is simplest to numerically compute the coefficients for higher particle numbers ($n > 2$). As eq.~\eqref{Eq:FFT2Clus} leads to a quite simple relation in terms of the new $C^\phi_{\textbf{2}^n,\lambda}$ coefficients, it is always best to start with solving the clustering equation at first. Then, the kinematic and bound state recursions (cf.~eqs.~\eqref{Eq:FFT2KP} and \eqref{Eq:FFT2BSP}) lead to a set of linear equations between the remaining coefficients that can be solved by numerical methods, e.g.~iteratively or by QR-decomposition.

Note that the solution depends on the continuous parameter $F^\phi_2/F^\phi_0$, which is not determined at this point. However, the finite number of scaling fields in the tricritical Ising model implies there can only be finite many allowed values of $F^\phi_2/F^\phi_0$. As shown in the next Subsection, it is determined by considering form factors containing odd particles (c.f.~also \cite{2022ScPP...12..162C}). Nevertheless, if $F^\phi_2/F^\phi_0$ and the physical normalisation for a given field $\phi$ are already known, the procedure outlined above can be used to generate the $A_2$ tower, from which all form factors containing only even particles can be computed using the bound state singularity equations. 

\subsection{Determining the ratio $F_2^\phi/F_0^\phi$}\label{subsec:f2f0ratio}

Next, we address the determination of the $F_2^\phi/F_0^\phi$ parameter. Although this was previously performed for $\Theta$ \cite{AMV}, and also $\mu$ and $\mu'$ \cite{2022ScPP...12..162C}, it is instructive to give a simple and independent derivation, which can also be useful for computing form factors of other fields in the $E_7$ model.

\subsubsection{Solution for the $\Theta$ field}

As stated above, determining $F_2^\Theta/F_0^\Theta$ requires going beyond the $A_2$ tower. It turns out to be sufficient to consider the $F_{11}^\Theta$ form factor, for which the Ansatz is
\begin{align}
    F_{11}^\Theta(\vartheta) &= \Tilde{Q}_{11}^\Theta(\vartheta)\frac{F^\text{min}_{11}(\vartheta)}{D_{11}(\vartheta)}\,,
\end{align}
where $\Tilde{Q}_{11}^\Theta(\vartheta)$ is an entire function of $\cosh\vartheta$, and we exploited Lorentz invariance by writing the form factor in terms of the rapidity difference $\vartheta = \vartheta_1 - \vartheta_2$. The bound on asymptotic growth only allows a $0$th order polynomial for $\Tilde{Q}_{11}^\Theta(\vartheta)$, which implies that it is a constant. Using the normalisation condition \cite{1991NuPhB.348..619Z,1993NuPhB.393..413F}
\begin{align}
    F_{aa}^\Theta(i\pi) = 2\pi m_a^2\,,
\end{align}
the result is
\begin{align} \label{Eq:F11Theta}
    F_{11}^\Theta(\vartheta) = 2\pi m_1^2\frac{F^\text{min}_{11}(\vartheta)}{D_{11}(\vartheta)}\,.
\end{align}
The form factor $F_2^\Theta$ can be computed by applying the bound state singularity equation \eqref{Eq:BSP} to the fusion $A_1\times A_1\rightarrow A_2$, which yields
\begin{align} \label{Eq:F2vsM1exact}
    \frac{F_2^\Theta\kappa_{22}^{-1/4}}{2\pi m_1^2} = \left(\Gamma_{11}^2\right)^{-1}\frac{2\sqrt{3}\sin^2\left(\frac{2\pi}{9}\right)}{\cos\frac{\pi}{9} + \sin\frac{\pi}{18}}\,,
\end{align}
where
\begin{align}
    \Gamma_{11}^2 = \sqrt{2\sqrt{3}\cot\frac{\pi}{18}\cot\frac{2\pi}{9}}\,.
\end{align}
We can express $F_2^\Theta/m_1^2$ numerically
\begin{align}
    F_2^\Theta/m_1^2 = 0.9604936853481771\ldots\,,
\end{align}
which agrees with the result in Ref.~\cite{AMV}.

The final step is to exploit the normalisation condition for the $A_2$ two-particle form factor. Using the Ansatz \eqref{Eq:FF2nAnsatz} for $F_{\textbf{2}^2}^\Theta$, the relation
\begin{align}
    F_{22}^\Theta(i\pi) &\equiv F_{\textbf{2}^2}^\Theta(\vartheta + i\pi, \vartheta) = 2\pi m_2^2
\end{align}
leads to
\begin{align}
    F_2^\Theta \kappa_{22}^{-1/4} C_{\textbf{2}^2,(2,1)}^\Theta &= 2\pi m_1^2 \left(2\cos\frac{5\pi}{18}\right)^2\,.
\end{align}
Utilizing eqs.~\eqref{Eq:C22v21} and \eqref{Eq:F2vsM1exact} we get a relation between the physical mass and the vacuum expectation value of the $\Theta$ field
\begin{align} \label{Eq:m1vsF0Theta}
    \frac{2\pi m_1^2}{F_0^\Theta} &= \frac{2\sqrt{3}\sin\frac{5\pi}{9}}{\sin\frac{2\pi}{9}}\,.
\end{align}
Multiplying the above formula with eq.~\eqref{Eq:F2vsM1exact}, we get an exact expression for $F_2^\Theta \kappa_{22}^{-1/4}/F_0^\Theta$, from which the $F_2^\Theta / F_0^\Theta$ ratio can be calculated numerically
\begin{align}
    F_2^\Theta / F_0^\Theta = 0.8113144869498665\ldots\,.
\end{align}

\subsubsection{Solution for the disorder fields} \label{sec:F2F0SL}

\begin{figure}
    \centering
    \includegraphics{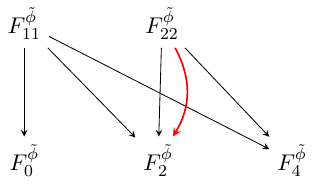}
    \caption{The form factor subsystem determining the ratio $F_2^{\Tilde{\phi}}/F_0^{\Tilde{\phi}}$ for $\Tilde{\phi}=\mu,\mu'$. The black arrows show kinematic and bound state singularity relations, while the red arrow corresponds to a relation given by the clustering property.}
    \label{fig:ff_subsystem}
\end{figure}

In the case of the disorder fields, our goal is to express every form factor coefficient in terms of the vacuum expectation value, $\langle\Tilde{\phi}\rangle = F_0^{\Tilde{\phi}}$ for $\Tilde{\phi}=\mu,\mu'$. To this end, we need a set of form factors for which the number of independent equations equals the number of free coefficients. The simplest such solvable subsystem  includes $F_0^{\Tilde{\phi}}, F_2^{\Tilde{\phi}}, F_4^{\Tilde{\phi}}, F_{11}^{\Tilde{\phi}}$ and $F_{22}^{\Tilde{\phi}}$ and is illustrated in Fig. \ref{fig:ff_subsystem}. It contains four bound state singularities:
\begin{align}
    \begin{aligned}
        \nonumber
        A_1\times A_1 &\rightarrow A_2\,,\\
        A_1\times A_1 &\rightarrow A_4\,,\\
        A_2\times A_2 &\rightarrow A_2\,,\\
        A_2\times A_2 &\rightarrow A_4\,.\\
    \end{aligned}
\end{align}
Additionally, $F_{11}^{\Tilde{\phi}}$ also has a kinematic pole relating $F_{11}^{\Tilde{\phi}}$ to $F_0^{\Tilde{\phi}}$ since $\Tilde{\phi}$ is semi-local respect to the $A_1$ particle. These relations lead to linear equations in terms of the coefficients of the $Q_{11}^{\Tilde{\phi}}$ and $Q_{22}^{\Tilde{\phi}}$ polynomials and reduce the space of solutions to a two-dimensional subspace. The last piece of information is supplied by the clustering property (denoted by a red arrow in Fig. \ref{fig:ff_subsystem}), which is quadratic and selects the two directions in this subspace corresponding to form factors of scaling fields. 

As a first step, we consider the $A_1$ two-particle form factor, for which semi-locality requires a modification of the Ansatz by including an extra denominator $\cosh\frac{\vartheta}{2}$ to account for its kinematic pole, and for the extra minus sign under cyclic property \cite{2022ScPP...12..162C}
\begin{align}
    F_{11}^{\Tilde{\phi}} (\vartheta) = \frac{1}{\cosh\frac{\vartheta}{2}} \Tilde{Q}_{11}^{\Tilde{\phi}}(\vartheta) \frac{F^\text{min}_{11}(\vartheta)}{D_{11}(\vartheta)}\,,
\end{align}
where $\vartheta = \vartheta_1 - \vartheta_2$ and $\Tilde{Q}_{11}^{\Tilde{\phi}}(\vartheta)$ is a polynomial in $\cosh\vartheta$, which the bound on the asymptotic growth restricts to first order. For later convenience, we rewrite the above Ansatz using notations similar to eq.~\eqref{Eq:FF2nAnsatz}
\begin{align} \label{Eq:F11SL}
    F^{\Tilde{\phi}}_{\textbf{1}^{2}}(\vartheta_1,\vartheta_{2}) = F_0^{\Tilde{\phi}}\frac{Q^{\Tilde{\phi}}_{\textbf{1}^{2}}(x_1,x_{2})}{(x_1x_{2})^{1/2}}\frac{F^\text{min}_{11}(\vartheta_1-\vartheta_2)}{D_{11}(\vartheta_1-\vartheta_2)(x_1+x_2)}\,,
\end{align}
where $x_k = e^{\vartheta_k}$ and we factored out a $F_0^{\Tilde{\phi}}$ normalisation factor. The homogeneous symmetric polynomial $Q^{\Tilde{\phi}}_{\textbf{1}^{2}}(x_1,x_2)$ is of second order polynomial in $x_1$ and $x_2$, and can be parameterised as
\begin{align} \label{Eq:Q11SL}
    Q^{\Tilde{\phi}}_{\textbf{1}^2}
    &= C^{\Tilde{\phi}}_{\textbf{1}^2,(2)}\sigma_2 + C^{\Tilde{\phi}}_{\textbf{1}^2,(1,1)}\sigma_1^2\,. 
\end{align}

The kinematic singularity equation must be carefully considered. When considering the form factor of the order field in the ferromagnetic phase, the $A_1$ particle is a kink that interpolates between the two vacua of opposite magnetisation; therefore, in the left-hand side of eq.~\eqref{Eq:KPE}, the form factor should be evaluated in the vacuum opposite to the one in which $F_0^{\phi}$ is determined, leaving to a relative minus sign. By Kramers-Wannier duality, this leads to a non-trivial and physically relevant minus sign for the disorder operator in the paramagnetic phase as well:
\begin{align}
    \begin{aligned} \label{Eq:F11KPE}
    -i\lim_{\Tilde{\vartheta}\rightarrow\vartheta} (\Tilde{\vartheta} - \vartheta) F^{\Tilde{\phi}}_{\textbf{1}^2}(\Tilde{\vartheta}+i\pi, \vartheta) &= -2 F_0^{\Tilde{\phi}}\,,
    \end{aligned}
\end{align}
and the equation leads to
\begin{align} \label{Eq:C11v2}
    C^{\Tilde{\phi}}_{\textbf{1}^2,2} &= 2\,.
\end{align}
The other free parameter of the $Q_{\textbf{1}^2}^{\Tilde{\phi}}$ is fixed by the $A_1\times A_1\rightarrow A_2$ fusion as
\begin{align} \label{Eq:C11v11}
    C^{\Tilde{\phi}}_{\textbf{1}^2,(1,1)} &= \left(2\cos\frac{5\pi}{18}\right)^{-2}\left[\frac{F_2^{\Tilde{\phi}}\kappa_{22}^{-1/4}}{F_0^{\Tilde{\phi}}}\Gamma_{11}^2\frac{\cos\frac{\pi}{9}+\sin\frac{\pi}{18}}{2\sqrt{3}\sin^2\left(\frac{2\pi}{9}\right)}2\cos\frac{5\pi}{18} - 2\right]\,.
\end{align}
At this point both $F_{11}^{\Tilde{\phi}}$ and $F_{22}^{\Tilde{\phi}}$ are expressed in terms of $F_{2}^{\Tilde{\phi}}/F_{0}^{\Tilde{\phi}}$ (c.f.~eqs.~\eqref{Eq:C22v111} and \eqref{Eq:C22v21}). The final equations to be utilized are the
\begin{align}
    \begin{aligned}
        \nonumber
        A_1\times A_1 &\rightarrow A_4\quad\text{and}\\
        A_2\times A_2 &\rightarrow A_4
    \end{aligned}
\end{align}
fusions. The equality of the $F_4^{\Tilde{\phi}}$ expressed using the two different fusions leads to a second order equation for $F_{2}^{\Tilde{\phi}}/F_{0}^{\Tilde{\phi}}$, which can be solved explicitly
\begin{align}
    \begin{aligned} \label{Eq:F2F0SLExact}
        \left(\frac{F_2^{\Tilde{\phi}}\kappa_{22}^{-1/4}}{F_0^{\Tilde{\phi}}}\right)_\pm =
        &-3^{1/4}\frac{2 \cos(\pi/18)^2 \csc(\pi/18)^{13/2}
  \sec(\pi/9) (-18 + 14 \cos(\pi/9) + 27 \sin(\pi/18))}{(\csc(\pi/36) - 
   \sec(\pi/36))^3 (\csc(\pi/36) + \sec(\pi/36))}\\
   &\pm\frac{3^{1/4}}{1 + 2 \sin(\pi/18)}\sqrt{\frac{-57 - 566 \cos(\pi/9) 
   + 611 \cos(2 \pi/9) + 695 \sin(\pi/18)}{71 - 72 \cos(\pi/9) + 16 \cos(2 \pi/9) - 
   90 \sin(\pi/18)}}\,.
    \end{aligned}
\end{align}
Evaluating the expression numerically, we find the same solutions as  Ref.~\cite{2022ScPP...12..162C}, which are associated with the two disorder operators as
\begin{align}
    F_{2}^{\mu}/F_{0}^{\mu} &= 0.3204131147841633\ldots\,,\\
    F_{2}^{\mu'}/F_{0}^{\mu'} &= 2.656500074781019\ldots\,.
\end{align}
Finally, the vacuum expectation values $F_{0}^{\mu}$ and $F_{0}^{\mu'}$ given in \eqref{eq:exact_muvevs} fixes the normalisation of the operators. The corresponding form factors of the order operators in the ferromagnetic phase can be obtained using Kramers--Wannier duality \eqref{eq:KW}.


\section{Systematic solution of the form factor bootstrap}\label{sec:A1}

As discussed above, the $\mathbb{Z}_2$ spin symmetry---distinguishing even and odd types of particles---prohibits the construction of form factors involving odd particles from the knowledge of the $A_2$ tower. An alternative approach to solve the form factor bootstrap is to determine the $A_1$ tower, i.e., form factors involving only the $A_1$ particle (an odd particle with the simplest pole structure). However, it comes with a price as the lack of self-fusion of $A_1$ (prohibited by the $\mathbb{Z}_2$ spin symmetry) complicates the structure of the recursive equations, as shown below. Nevertheless, it is possible to find a system of recursive equations which allows the complete determination of the $A_1$ tower, leading to all the form factors.

The kinematic and bound state singularity equations connect multi-particle states with the same parity under the $\mathbb{Z}_2$ spin symmetry. Therefore, we first focus on the even sector, i.e., form factors involving an even number of $A_1$ particles. As discussed in Subsection \ref{subsec:f2f0ratio}, the Ansatz for the $\Theta$ and the semi-local fields differ for the $A_1$ form factors, so we treat them separately. For the $\Theta$ field, the $A_1$ tower form factor Ansatz is a suitable generalisation of eq.~\eqref{Eq:F11Theta} as
\begin{align}
    F^\Theta_{\textbf{1}^{2n}}(\vartheta_1, \ldots, \vartheta_{2n}) = H^\Theta_{\textbf{1}^{2n}}\frac{Q^\Theta_{\textbf{1}^{2n}}(x_1,\ldots,x_{2n})}{(x_1\cdot\ldots\cdot x_{2n})^{n-1}}\prod_{k<l}^{2n}\frac{F^\text{min}_{11}(\vartheta_k-\vartheta_l)}{D_{11}(\vartheta_k-\vartheta_l)(x_k+x_l)}\,,
\end{align}
where
\begin{align}
    H^\Theta_{\textbf{1}^{2n}} &= 2\pi m_1^2 \kappa_{11}^{-n(n-1)}\,,
\end{align}
and with the boundaries on the $Q^\Theta_{\textbf{1}^{2n}}$ as
\begin{align}
    Q^\Theta_{\textbf{1}^{2}} &= \sigma_1\,,\nonumber\\
    \deg_1\left(Q^\Theta_{\textbf{1}^{2n}}\right) &\leq 4n - 3\,,\nonumber\\
    \deg_2\left(Q^\Theta_{\textbf{1}^{2n}}\right) &\leq 8n - 7\,,\nonumber\\
    &\,\,\,\vdots\nonumber\\
    \label{Eq:partDegTh}
    \deg_k\left(Q^{\Tilde{\phi}}_{\textbf{1}^{2n}}\right) &\leq \left\lfloor k\left(4n-k-\frac{3}{2}\right)\right\rfloor\,,\\
    &\,\,\,\vdots\nonumber\\
    \deg\left(Q^\Theta_{\textbf{1}^{2n}}\right) &= 4n^2-3n\,.\nonumber
\end{align}
For the disorder fields, the Ansatz must be modified to take into account their semi-locality phase as in eq.~\eqref{Eq:F11SL}:
\begin{align} \label{Eq:SLAnsatz}
    F^{\Tilde{\phi}}_{\textbf{1}^{2n}}(\vartheta_1,\ldots,\vartheta_{2n}) = H^{\Tilde{\phi}}_{\textbf{1}^{2n}}\frac{Q^{\Tilde{\phi}}_{\textbf{1}^{2n}}(x_1,\ldots,x_{2n})}{(x_1\cdot\ldots\cdot x_{2n})^{n-1/2}}\prod_{k<l}^{2n}\frac{F^\text{min}_{11}(\vartheta_k-\vartheta_l)}{D_{11}(\vartheta_k-\vartheta_l)(x_k+x_l)}\,,
\end{align}
where
\begin{align}
    H^{\Tilde{\phi}}_{\textbf{1}^{2n}} &= F^{\Tilde{\phi}}_0 \kappa_{11}^{-n(n-1)}\,,
\end{align}
and the restrictions concerning the $Q^{\Tilde{\phi}}_{\textbf{1}^{2n}}$ polynomials is as follows
\begin{align}
    Q^{\Tilde{\phi}}_{\textbf{1}^{0}} &= -1\,,\nonumber\\
    \deg_1\left(Q^{\Tilde{\phi}}_{\textbf{1}^{2n}}\right) &\leq 4n - 2\,,\nonumber\\
    \deg_2\left(Q^{\Tilde{\phi}}_{\textbf{1}^{2n}}\right) &\leq 8n - 6\,,\nonumber\\
    &\,\,\,\vdots\nonumber\\
    \label{Eq:partDegSl}
    \deg_k\left(Q^{\Tilde{\phi}}_{\textbf{1}^{2n}}\right) &\leq k(4n-k-1)\,,\\
    &\,\,\,\vdots\nonumber\\
    \deg\left(Q^{\Tilde{\phi}}_{\textbf{1}^{2n}}\right) &= 4n^2-2n\nonumber\,.
\end{align}
Note that $Q^{\Tilde{\phi}}_{\textbf{1}^0}$ is chosen such that $F^{\Tilde{\phi}}_{\textbf{1}^0} = -F^{\Tilde{\phi}}_0$ in alignment with eq.~\eqref{Eq:F11KPE}.

We remark that---similarly to the case of the $A_2$ tower---the correct choice of the numerators $(x_1\cdot\ldots\cdot x_{2n})^{n-1}$ and $(x_1\cdot\ldots\cdot x_{2n})^{n-1/2}$ is not immediately apparent. They can be obtained from eqs.~\eqref{Eq:F11Theta} and \eqref{Eq:F11SL} for $2n = 2$ case, and then their $n$-dependence is fixed by the kinematic and bound state singularity equations.

\subsection{Three-$A_1$ fusion equations}

The normalization is set so that the initial $Q$ polynomials have a simple form. The $Q$ polynomials corresponding to a higher number of particles are determined recursively from the lower ones via kinematic and bound state singularity equations.

\begin{figure}[t]
\begin{center}
    \includegraphics{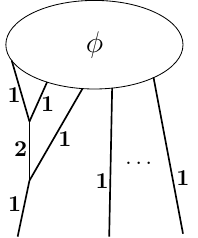}
    ~~~~~~~~~~~~~~~~~~
    \includegraphics{FC111to1via2.pdf}
\end{center}
    \caption{Schematic picture of fusion chains \eqref{Eq:111to1}.}
    \label{fig:111to1}
\end{figure}

The derivation of the kinematic singularity equation is straightforward. For the $\Theta$ field, we obtain
\begin{align}
    \begin{aligned} \label{Eq:KPTheta}
        Q^\Theta_{\textbf{1}^{2(n+1)}}(-x,x,x_1,\ldots,x_{2n}) &= -i x U^\Theta_{\textbf{1}^{2n}}(x|\{x_k\}) Q^\Theta_{\textbf{1}^{2n}}(x_1,\ldots,x_{2n})\,,
    \end{aligned}
\end{align}
where
\begin{align}
    \begin{aligned}
        U^\Theta_{\textbf{1}^{2n}}(x|\{x_k\}) &= \prod_{k=1}^{2n}\prod_{\alpha\in\mathcal{A}_{11}}[\alpha]_k[-\Bar{\alpha}]_k - \prod_{k=1}^{2n}\prod_{\alpha\in\mathcal{A}_{11}}[-\alpha]_k[\Bar{\alpha}]_k\,.
    \end{aligned}
\end{align}
For the disorder fields it reads as
\begin{align}
    \begin{aligned} \label{Eq:KPSL}
        Q^{\Tilde{\phi}}_{\textbf{1}^{2(n+1)}}(-x,x,x_1,\ldots,x_{2n}) &= x^2 U^{\Tilde{\phi}}_{\textbf{1}^{2n}}(x|\{x_k\}) Q^{\Tilde{\phi}}_{\textbf{1}^{2n}}(x_1,\ldots,x_{2n})\,,
    \end{aligned}
\end{align}
where
\begin{align}
    \begin{aligned}
        U^{\Tilde{\phi}}_{\textbf{1}^{2n}}(x|\{x_k\}) &= \prod_{k=1}^{2n}\prod_{\alpha\in\mathcal{A}_{11}}[\alpha]_k[-\Bar{\alpha}]_k + \prod_{k=1}^{2n}\prod_{\alpha\in\mathcal{A}_{11}}[-\alpha]_k[\Bar{\alpha}]_k\,.
    \end{aligned}
\end{align}
Extracting the information encoded in the bound state equations is more complicated. Since the $A_1$ particle does not exhibit self-fusion, the simplest recursive equation involves two consecutive fusions
\begin{align}
    \begin{aligned} \label{Eq:111to1}
        (A_1\times A_1)&\times A_1 \rightarrow A_2 \times A_1 \rightarrow A_1\quad\text{or}\\
        (A_1\times A_1)&\times A_1 \rightarrow A_4 \times A_1 \rightarrow A_1\quad\,,
    \end{aligned}
\end{align}
that we also illustrated in Fig.~\ref{fig:111to1}.
In both cases, however, the fusion angles are such that all these relations are implied by the kinematic singularity equation and, therefore, redundant.

\begin{figure}[t]
    \centering
    \includegraphics{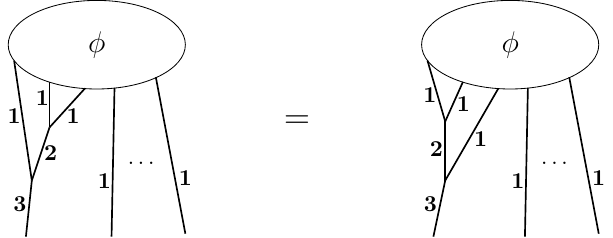}
    \caption{Schematic illustration of fusion chain \eqref{Eq:FC3to3}.}
    \label{fig:111to111}
\end{figure}

To obtain effective constraints from the bound state structure, it turns out to be necessary to combine four bound state singularity equations\footnote{A three-fold fusion starting from $A_1$ type particles always results in an even type particle and thus leads out of the $A_1$ tower.}. A suitable way is to utilize the following fusion chain
\begin{align}
    \label{Eq:FC3to3}
    A_1\times (A_1\times A_1)\rightarrow A_1\times A_2\rightarrow A_3\leftarrow A_2\times A_1 \leftarrow (A_1\times A_1)\times A_1\,,
\end{align}
which indicates that $A_3$ can be obtained by fusing $3$ $A_1$ particles in two independent ways, which is also illustrated in Fig.~\ref{fig:111to111}.
For the $\Theta$ field, the above fusion chain gives the following equation in terms of the $Q_{\textbf{1}^{2n}}^\Theta$ polynomial
\begin{align}
    \begin{aligned} \label{Eq:BSPTheta}
        &e^{\frac{i\pi}{18}(8n-9)}\prod_{k=4}^{2n}\left(\left[\frac{12}{18}\right]_k\left[-\frac{16}{18}\right]_k\right)\, Q^{\Theta}_{\textbf{1}^{2n}}\left(e^{\frac{4}{18}i\pi}x, e^{\frac{2}{18}i\pi}x, e^{-\frac{8}{18}i\pi}x, x_4,\ldots,x_{2n}\right)=\\
        &= e^{-\frac{i\pi}{18}(8n-9)}\prod_{k=4}^{2n}\left(\left[-\frac{12}{18}\right]_k\left[\frac{16}{18}\right]_k\right)\, Q^{\Theta}_{\textbf{1}^{2n}}\left(e^{\frac{8}{18}i\pi}x, e^{-\frac{2}{18}i\pi}x, e^{-\frac{4}{18}i\pi}x, x_4,\ldots,x_{2n}\right)\,,
    \end{aligned}
\end{align}
while for the semi-local fields, it results in the equation
\begin{align}
    \begin{aligned} \label{Eq:BSPSL}
        &e^{\frac{8}{18}i\pi(n-1)}\prod_{k=4}^{2n}\left(\left[\frac{12}{18}\right]_k\left[-\frac{16}{18}\right]_k\right)\, Q^{\Tilde{\phi}}_{\textbf{1}^{2n}}\left(e^{\frac{4}{18}i\pi}x, e^{\frac{2}{18}i\pi}x, e^{-\frac{8}{18}i\pi}x, x_4,\ldots,x_{2n}\right)=\\
        &= e^{-\frac{8}{18}i\pi(n-1)}\prod_{k=4}^{2n}\left(\left[-\frac{12}{18}\right]_k\left[\frac{16}{18}\right]_k\right)\, Q^{\Tilde{\phi}}_{\textbf{1}^{2n}}\left(e^{\frac{8}{18}i\pi}x, e^{-\frac{2}{18}i\pi}x, e^{-\frac{4}{18}i\pi}x, x_4,\ldots,x_{2n}\right)\,.
    \end{aligned}
\end{align}
These equations are not recursions as they relate the form factor $Q^\phi_{\textbf{1}^{2n}}$ to itself. Rather, these three-$A_1$ fusion relations give linear equations in terms of the coefficients of $Q^\phi_{\textbf{1}^{2n}}$, which further reduce the space of possible solutions.

We claim that the above fusion equation, along with the kinematic singularity equation, is enough to recursively determine all form factors (of the $\Theta$ and the semi-local fields) in the $A_1$ tower when $2n \geq 6$. For $2n = 6$, this statement can be checked by direct analytic computation, while for $2n > 6$, we give a mathematical proof in Appendix \ref{sec:proof}. Additional information is necessary for $2n = 4$, as discussed in the following.

\subsection{Solving the $A_1$ tower}

\subsubsection{The $\Theta$ field}

For the $\Theta$ field form factors there are additional constrains on the $Q_{\textbf{1}^{2n}}^\phi$ polynomials. The $\partial^\mu T_{\mu\nu}(x)$ conservation law of the energy-momentum tensor implies that the form factors of $\Theta(x)$ contains the Lorentz-invariant momentum square $P^2 = (p_1 + \ldots p_{2n})^2$, which can be written in terms of symmetric polynomials of $x_k = e^{\vartheta_k}$ as
\begin{align}
    P^2 = m_1^2\frac{\sigma_1\sigma_{2n-1}}{\sigma_{2n}}\,.
\end{align}
Since $F_{\textbf{1}^{2n}}^\Theta$ contains $P^2$, the $Q_{\textbf{1}^{2n}}^\Theta$ must have the form
\begin{align} \label{Eq:P2QFactorTheta}
    Q_{\textbf{1}^{2n}}^\Theta = \sigma_1\sigma_{2n-1}\hat{Q}_{\textbf{1}^{2n}}^\Theta
\end{align}
for $2n > 2$. The remaining task is to determine the polynomials $\hat{Q}_{\textbf{1}^{2n}}^\Theta$ polynomials, which have lower partial degrees than those given in \eqref{Eq:partDegSl}:
\begin{align}
    \deg_k\left(\hat{Q}_{\textbf{1}^{2n}}^\Theta\right) \leq \left\lfloor k\left(4n-k-\frac{5}{2}\right)\right\rfloor - 1
\end{align}
and
\begin{align}
    \deg_{2n}\left(\hat{Q}_{\textbf{1}^{2n}}^\Theta\right) &= 4n^2 - 5n\,.
\end{align}
This can also be utilized in the $A_2$ tower calculation to reduce the number of coefficients and simplify the calculations significantly.

For $2n = 2$, the $Q_{\textbf{1}^{2n}}^\Theta$ polynomial is already determined (c.f.~eq.~\eqref{Eq:F11Theta})
\begin{align}
    Q_{\textbf{1}^{2}}^\Theta(x_1, x_2) &= \sigma_1(x_1, x_2) \equiv x_1 + x_2\,.
\end{align}
As claimed, eqs.~\eqref{Eq:KPTheta} and \eqref{Eq:BSPTheta} fully determine the $Q_{\textbf{1}^{2n}}^\Theta$ polynomials when $2n \geq 6$. Thanks to the restriction \eqref{Eq:P2QFactorTheta}, this is also true for $2n = 4$. In summary, all the form factors of the $\Theta$ field in the $A_1$ tower can be determined as a systematic solution of the recursion relations.

\subsubsection{The semi-local fields}

For $2n = 2$, the $Q^{\Tilde{\phi}}_{\textbf{1}^{2}}$ polynomial is given in Section \ref{sec:F2F0SL} (c.f.~eqs.~\eqref{Eq:Q11SL}, \eqref{Eq:C11v2} and \eqref{Eq:C11v11}). Note that the $C^{\Tilde{\phi}}_{\textbf{1}^2,(1,1)}$ coefficient depends on the ratio $F_2^{\Tilde{\phi}}\kappa_{22}^{-1/4}/F_0^{\Tilde{\phi}}$ that was already determined in eq.~\eqref{Eq:F2F0SLExact}.

For $2n = 4$, the $Q^{\Tilde{\phi}}_{\textbf{1}^{4}}$ polynomial has 16 coefficients. However, the combined rank of the system of linear equations coming from eqs.~\eqref{Eq:KPSL} and \eqref{Eq:BSPSL} is only $15$. So, the kinematic and bound state singularity equation has a one-dimensional kernel. We utilize the formerly determined $A_2$ tower to fix this ambiguity by fusing all the $A_1$'s to the $A_2$'s pairwise. We get the following equation in terms of the $Q^{\Tilde{\phi}}_{\textbf{1}^{2n}}$ and $Q^{\Tilde{\phi}}_{\textbf{2}^{n}}$ polynomials
\begin{align}
    \begin{aligned} \label{Eq:A1toA2SL}
        Q^{\Tilde{\phi}}_{\textbf{1}^{2n}}(\varphi x_1,\varphi^{-1}x_1,&\ldots,\varphi x_n,\varphi^{-1}x_n) =\\  &\frac{F_2^{\Tilde{\phi}}\kappa_{22}^{-1/4}}{F_0^{\Tilde{\phi}}}\left(\Gamma_{11}^2\frac{\cos\frac{\pi}{9}+\sin\frac{\pi}{18}}{2\sqrt{3}\sin^2\left(\frac{2\pi}{9}\right)}\right)^{n}\left(2\cos\frac{5\pi}{18}\right)^n\left(\prod_{k=1}^n x_k\right)^2\cdot\\
        &\cdot\prod_{k<l}^n\left(-\left[\Bar{0}\right]_{kl}\left[\frac{10}{18}\right]_{kl}\left[-\frac{10}{18}\right]_{kl}\left[\Bar{\frac{2}{18}}\right]_{kl}\left[-\Bar{\frac{2}{18}}\right]_{kl}\right)Q^{\Tilde{\phi}}_{\textbf{2}^{n}}(x_1,\ldots, x_n)\,,
    \end{aligned}
\end{align}
where $\varphi = e^{5i\pi/18}$ and we introduced the notations
\begin{align}
    [\alpha]_{kl} = x_k - e^{i\pi\alpha}x_l\quad\text{and} \quad [\bar{\alpha}]_{kl} = x_k + e^{i\pi\alpha}x_l\,.
\end{align}
Combined with the kinematic and bound state singularity equation, this relation completely fixes the 4-particle $A_1$ form factor. For $2n \geq 6$, the above equation is not needed to fix the form factor but is useful as a crosscheck.

The $A_1$ tower can also be determined without relying on the $A_2$ tower, which we used above for the $2$ and $4$-particle form factors. Then the $C^{\Tilde{\phi}}_{\textbf{1}^2,(1,1)}$ coefficient is a free parameter, and there appears another free parameter at the level of four particles. These free parameters can be fixed from the $A_1^4 \rightarrow A_1^2\times A_1^2$ clustering equation, which determines the free parameter in the 4-particle form factor and gives a second order equation for $C^{\Tilde{\phi}}_{\textbf{1}^2,(1,1)}$. The two solutions for $C^{\Tilde{\phi}}_{\textbf{1}^2,(1,1)}$ corresponds to the two disorder operators. This procedure is analogous to our calculation, with the difference that we used the clustering inside the $A_2$ tower and traced back the information via the $A_1^{2n}\rightarrow A_2^n$ fusions, as illustrated with the commutative diagram below. So, in principle, the $A_1$ tower can be fully determined independently. However, the calculation is more transparent if the $A_2$ tower is also used.
\begin{center}
\begin{tikzpicture}[>=stealth, node distance=1.5cm]
    \node (F1111) at (0,3) {$F^{\Tilde{\phi}}_{1111}$};
    \node[right=of F1111] (F22) {$F^{\Tilde{\phi}}_{22}$};
    \node[below=of F1111] (F11) {$\,\,F^{\Tilde{\phi}}_{11}\,$};
    \node[right=of F11] (F2) {$\,F^{\Tilde{\phi}}_{2}\,$};
    
    \draw[->] (F1111) -- (F22) node[midway, above] {Fusion};
    \draw[->] (F11) -- (F2) node[midway, below] {Fusion};
    \draw[->] (F1111) -- (F11) node[midway, left] {Clustering};
    \draw[->] (F22) -- (F2) node[midway, right] {Clustering};
\end{tikzpicture}
\end{center}

\subsubsection{Parity invariance}

An easy method to eliminate many unknown coefficients is to exploit parity ($\mathbb{P}$) invariance $x\rightarrow -x$. The asymptotic states transform under spatial reflection as
\begin{align}
    \mathbb{P}|A_{a_1}(\vartheta_1)\ldots A_{a_n}(\vartheta_n)\rangle &= |A_{a_n}(-\vartheta_n)\ldots A_{a_1}(-\vartheta_1)\rangle\,,
\end{align}
and so the identity
\begin{align}
    F_{a_1,\ldots,a_n}^\phi(\vartheta_1,\ldots,\vartheta_n) &= F_{a_n,\ldots,a_1}^\phi(-\vartheta_n,\ldots,-\vartheta_1)
\end{align}
holds for form factors of all parity-invariant fields such as $\Theta$ and the order/disorder operators. In terms of the $Q$ polynomials, the above equation leads to the following relations
\begin{align}
    \left(x_1\cdot\ldots\cdot x_{2n}\right)^{4n-3} Q^\Theta_{\textbf{1}^{2n}}(x_1^{-1},\ldots,x_{2n}^{-1}) &= Q^\Theta_{\textbf{1}^{2n}}(x_1,\ldots,x_{2n})
\end{align}
and
\begin{align}
    \left(x_1\cdot\ldots\cdot x_{2n}\right)^{4n-2} Q^{\Tilde{\phi}}_{\textbf{1}^{2n}}(x_1^{-1},\ldots,x_{2n}^{-1}) &= Q^{\Tilde{\phi}}_{\textbf{1}^{2n}}(x_1,\ldots,x_{2n})
\end{align}
for $\Theta$ and the semi-local fields, respectively. 

The solution of the recursive equation for a generic $n$ is as follows. First, one determines all the different coefficients of $\hat{Q}_{\textbf{1}^{2n}}^\Theta$ or $Q_{\textbf{1}^{2n}}^{\Tilde{\phi}}$ allowed by the partial and total degree restrictions (similarly to eqs.~\eqref{Eq:Q2nLinComb}, \eqref{Eq:partToDegTotA2n}, and  \eqref{Eq:partToDegkA2n}). Then, applying parity invariance eliminates a large number of unknowns. Table \ref{tab:PinvCoeffs} illustrates the number of free coefficients of the $\hat{Q}_{\textbf{1}^{2n}}^\Theta$ and the $Q_{\textbf{1}^{2n}}^{\Tilde{\phi}}$ polynomials before and after applying parity invariance. Finally, the kinematic and bound state singularity equations are exploited, yielding a system of linear equations in terms of the coefficients of the $Q_{\textbf{1}^{2n}}^\phi$ polynomial. For form factors including several particles, the number of coefficients increases drastically, and their symbolic computation is prohibitively expensive. Therefore, we used QR-decomposition to numerically solve the over-determined linear equations derived from the kinematic and bound state singularity equations. 

\begin{table}[h!]
    \centering
    \renewcommand*{\arraystretch}{1.5}
    \begin{tabular}{|c|c|c|c|c|} \hline
    $2n$ & $\#\left(\hat{Q}_{\textbf{1}^{2n}}^\Theta\right)$ & $\#\left(\hat{Q}_{\textbf{1}^{2n}}^\Theta\right)$ after $\mathbb{P}$-inv. & $\#\left(Q_{\textbf{1}^{2n}}^{\Tilde{\phi}}\right)$ & $\#\left(Q_{\textbf{1}^{2n}}^{\Tilde{\phi}}\right)$ after $\mathbb{P}$-inv. \\ \hline \hline
    4 & 5 & 4 & 16 & 12 \\ \hline
    6 & 88 & 52 & 247 & 143 \\ \hline
    8 & 2100 & 1099 & 5302 & 2756 \\ \hline
    
    \end{tabular}
    \caption{Number of free coefficients of the $Q_{\textbf{1}^{2n}}^{\phi}$ polynomials (denoted by $\#\left(\hat{Q}_{\textbf{1}^{2n}}^\Theta\right)$) before and after applying parity invariance.}
    \label{tab:PinvCoeffs}
\end{table}

We constructed the $A_1$ tower up to $8$ particles. The numerical problem is ill-conditioned, especially for the higher form factors with many coefficients, which can be solved by brute force arbitrary precision arithmetic.

\subsection{Odd elements of the $A_1$ tower}

Form factors corresponding to multi-particle states with odd $\mathbb{Z}_2$ parity have non-vanishing matrix elements with $\sigma, \sigma'$ in the paramagnetic phase (and correspondingly $\mu,\mu'$ in the ferromagnetic phase). In this subsection, we denote with $\phi$ the field dual to the semi-local field $\Tilde{\phi}$.

The Ansatz for the odd elements of the $A_1$ tower is identical to eq.~\eqref{Eq:SLAnsatz} with the $2n\rightarrow 2n+1$ substitution
\begin{align}
    F^\phi_{\textbf{1}^{2n+1}}(\vartheta_1,\ldots,\vartheta_{2n+1}) = H^\phi_{\textbf{1}^{2n+1}}\frac{Q^\phi_{\textbf{1}^{2n+1}}(x_1,\ldots,x_{2n+1})}{(x_1\cdot\ldots\cdot x_{2n})^{n}}\prod_{k<l}^{2n+1}\frac{F^\text{min}_{11}(\vartheta_k-\vartheta_l)}{D_{11}(\vartheta_k-\vartheta_l)(x_k+x_l)}\,,
\end{align}
where
\begin{align}
    H^\phi_{\textbf{1}^{2n+1}} = F_0^{\Tilde{\phi}}\kappa_{11}^{-n^2+1/4}\,,
\end{align}
and the constraints on the $Q^\phi_{\textbf{1}^{2n+1}}$ polynomials are as follows
\begin{align}
    Q^\phi_{\textbf{1}^{1}} &= \frac{F_1^\phi\kappa_{11}^{-1/4}}{F_0^{\Tilde{\phi}}}\,,\\
    \deg_k\left(Q^\phi_{\textbf{1}^{2n+1}}\right) &\leq k(4n-k+1)\,,\\
    \deg\left(Q^\phi_{\textbf{1}^{2n+1}}\right) &= 4n^2 + 2n\,.
\end{align}
To determine the odd elements of the $A_1$ tower, we follow \cite{2022ScPP...12..162C} 
and exploit the clustering property in the form relating form factors of different fields with the same conformal weight (e.g., $\sigma \leftrightarrow \mu$ and $\sigma'\leftrightarrow\mu'$), which allows to determine the odd elements of the $A_1$ tower in terms of the even ones. The $A_1^{2n}\rightarrow A_1\times A_1^{2n-1}$ clustering gives
\begin{align} \label{Eq:ClusSLtoLoc}
    \lim_{X\rightarrow \infty}Q^{\Tilde{\phi}}_{\textbf{1}^{2n}}(X,x_2,\ldots,x_{2n})X^{-(4n-2)} &= -(-1)^{n}\frac{F_1^\phi\kappa_{11}^{-1/4}}{F_0^{\Tilde{\phi}}} Q^{\phi}_{\textbf{1}^{2n-1}}(x_2,\ldots,x_{2n})\,.
\end{align}
For $n = 1$, we obtain the $F_1^\phi/F_0^{\Tilde{\phi}}$ ratio
\begin{align}
    \frac{F_1^\phi\kappa_{11}^{-1/4}}{F_0^{\Tilde{\phi}}} &= \sqrt{C^{\Tilde{\phi}}_{\textbf{1}^2,(1,1)}}\,,
\end{align}
where $C^{\Tilde{\phi}}_{\textbf{1}^2,(1,1)}$ is already determined in eq.~\eqref{Eq:C11v11} and $F_1^\phi/F_0^{\Tilde{\phi}}$ is chosen to be positive real. Evaluating the ratio numerically, we get
\begin{align}
    F_1^\sigma/F_0^\mu &= 0.4920045700848942\ldots\,,\\
    F_1^{\sigma'}/F_0^{\mu'} &= 2.6624700017785751\ldots
\end{align}
in the paramagnetic phase, which agrees with the results in Ref.~\cite{2022ScPP...12..162C}. Once $F_1^\phi/F_0^{\Tilde{\phi}}$ is known, eq.~\eqref{Eq:ClusSLtoLoc} systematically determines the odd elements of the $A_1$ tower from the even ones.

We remark that the clustering property can be formulated in many different settings of the asymptotic limit $A_1^n\rightarrow A_1^k\times A_1^{n-k}$. One notable version is the companion of eq.~\eqref{Eq:ClusSLtoLoc} where the even elements are determined from the odd ones:
\begin{align}
    \lim_{X\rightarrow \infty}Q^{\phi}_{\textbf{1}^{2n+1}}(X,x_1,\ldots,x_{2n})X^{-4n} &= -(-1)^{n}\frac{F_1^\phi\kappa_{11}^{-1/4}}{F_0^{\Tilde{\phi}}} Q^{\Tilde{\phi}}_{\textbf{1}^{2n}}(x_1,\ldots,x_{2n})\,.
\end{align}
This relation is not actually necessary to calculate the $A_1$ tower but serves as a good crosscheck.

We further remark that going one particle higher ($2n$) might be wasteful to determine an odd state ($2n-1$) form factor. The odd subtower can also be systematically determined on its own by using eqs.~\eqref{Eq:KPSL} and \eqref{Eq:BSPSL} with the $2n\rightarrow 2n + 1$ substitution. In this case, the kinematic and bound state pole equations fully determine $Q^\phi_{\textbf{1}^{2n+1}}$ whenever $2n+1 \geq 5$ (the $2n + 1 = 5$ case can be checked manually, while the proof in Appendix \ref{sec:proof} works the same way for $2n + 1 \geq 7$). The $2n+1 = 3$ case should be handled separately where we need to exploit some extra information using the clustering property.


\section{Crosschecks and general form factors from the $A_1$ tower}\label{sec:crosschecks}

Our results can be crosschecked by comparing to existing ones in the literature by constructing certain specific form factors from the $A_1$ tower. We also describe an efficient method suitable for constructing general multi-particle form factors.

The $A_1$ is a generating particle: every particle can be constructed from it via fusion (i.e., via the consecutive application of eq.~\eqref{Eq:BSP}). Table \ref{tab:A1weights} summarises how many $A_1$ particles are necessary to fuse into a given particle. Note that $F_{\textbf{1}^8}^\phi$ is already sufficient to deduce all two-particle form factors.

\begin{table}[h!]
    \centering
    \begin{tabular}{|c||c|c|c|c|c|c|c|} \hline
        $A_a$ & $A_1$ & $A_2$ & $A_3$ & $A_4$ & $A_5$ & $A_6$ & $A_7$ \\ \hline
        $A_1$ weight & 1 & 2 & 3 & 2 & 4 & 3 & 4 \\ \hline
    \end{tabular}
    \caption{Number of $A_1$ particles necessary to fuse into an $A_a$ particle.}
    \label{tab:A1weights}
\end{table}

\subsection{Crosschecks for the $\Theta$ field}

First, we consider the $\Theta$ field form factors. Form factors of odd multi-particle states are identically zero, so we have yet to consider only the even ones. The one particle form factors can be deduced from $F_{\textbf{1}^2}^\Theta$ and $F_{\textbf{1}^4}^\Theta$. 
For the two-particle form factors, we adopt the parameterisation analogous to Ref.~\cite{AMV}
\begin{align}
    F_{ab}^{\Theta}(\vartheta) &= m_1^2\frac{F_{ab}^\text{min}(\vartheta)}{D_{ab}(\vartheta)}\left(\cosh\vartheta + \frac{m_a^2+m_b^2}{2m_am_b}\right)^{1-\delta_{ab}}\sum_{k=0}^{N^\Theta_{ab}} a_{ab}^k\cosh^k(\vartheta)\,.
\end{align}
The maximum exponent $N_{ab}^\Theta$ is 
\begin{align}
    N_{ab}^\Theta = \left\lfloor C_{ab}^{-1} \right\rfloor-1\,,
\end{align}
where
\begin{align}
    C_{ab}^{-1} &= \frac{1}{2}\sum_{\alpha\in\overline{\mathcal{A}_{ab}}} p_{ab}(\alpha)
\end{align}
is the inverse of the $E_7$ Cartan matrix \cite{1997hep.th....5142P}. The results for the one and the lowest few two-particle form factors are summarised in Table \ref{tab:FFThetaNumeric}, with the $F_{33}^\Theta$ and $F_{25}^\Theta$ computed from the six-particle form factor $F_{\textbf{1}^6}^\Theta$. These results agree perfectly with Ref.~\cite{AMV}.

\begin{table}[h!]
    \centering
    \renewcommand*{\arraystretch}{1.2}
    \begin{tabular}{||c|c||c|c||} \hline
        $F_2^{\Theta}/m_1^2$ & $0.9604936853$ & $a_{11}^0$ & $6.283185307$ \\ \hline
        $F_4^{\Theta}/m_1^2$ & $-0.4500141924$ & $a_{22}^0$ & $15.09207695$ \\ \hline
        $F_5^{\Theta}/m_1^2$ & $0.2641467198$ & $a_{22}^1$ & $4.707833688$ \\ \hline
        $F_7^{\Theta}/m_1^2$ & $-0.05569063847$ & $a_{13}^0$ & $30.70767637$ \\ \hline
        \multicolumn{2}{c|}{} & $a_{24}^0$ & $79.32168251$ \\ \cline{3-4}
        \multicolumn{2}{c|}{} & $a_{24}^1$ & $16.15028003$ \\ \cline{3-4}
        \multicolumn{2}{c|}{} & $a_{33}^0$ & $295.3281130$ \\ \cline{3-4}
        \multicolumn{2}{c|}{} & $a_{33}^1$ & $396.9648559$ \\ \cline{3-4}
        \multicolumn{2}{c|}{} & $a_{33}^2$ & $123.8295119$ \\ \cline{3-4}
        \multicolumn{2}{c|}{} & $a_{25}^0$ & $-3534.798444$ \\ \cline{3-4}
        \multicolumn{2}{c|}{} & $a_{25}^1$ & $-4062.255130$ \\ \cline{3-4}
        \multicolumn{2}{c|}{} & $a_{25}^2$ & $-556.5589101$ \\ \cline{3-4}
    \end{tabular}
    \caption{One-particle form factors and two-particle form factor coefficients of the $\Theta$ field.}
    \label{tab:FFThetaNumeric}
\end{table}

\subsection{Crosschecks for the order and disorder fields}

We consider the form factors of the $\mu, \mu'$ and the $\sigma, \sigma'$ fields in the paramagnetic phase, where the $\sigma$ and $\sigma'$ are local fields coupling with odd multi-particle states, while the $\mu$ and $\mu'$ are semi-local fields and coupling with even multi-particle states. 

The one-particle form factors of $\mu$ and $\mu'$ can be computed from $F_{\textbf{1}^2}^{\Tilde{\phi}}$ and $F_{\textbf{1}^4}^{\Tilde{\phi}}$, while the one-particle form factors of $\sigma$ and $\sigma'$ are constructed from $F_{\textbf{1}^3}^{\Tilde{\phi}}$, with the results summarised in Table \ref{tab:1partFFs}.

\begin{table}[h!]
    \centering
    \renewcommand*{\arraystretch}{1.2}
    \begin{tabular}{||c|c|c||c|c|c||} \hline
        $F_a^{\Tilde{\phi}}/F_0^{\Tilde{\phi}}$ & $\mu$ & $\mu'$ & $F_a^{\phi}/F_0^{\Tilde{\phi}}$ & $\sigma$ & $\sigma'$ \\ \hline \hline
        $F_2^{\Tilde{\phi}}/F_0^{\Tilde{\phi}}$ & $0.3204131148$ & $2.656500075$ & $F_1^{\phi}/F_0^{\Tilde{\phi}}$ & $0.4920045701$ & $2.662470002$ \\ \hline
        $F_4^{\Tilde{\phi}}/F_0^{\Tilde{\phi}}$ & $-0.1320143535$ & $-1.808904698$ & $F_3^{\phi}/F_0^{\Tilde{\phi}}$ & $-0.1747403347$ & $-2.219610007$ \\ \hline
        $F_5^{\Tilde{\phi}}/F_0^{\Tilde{\phi}}$ & $0.06960044886$ & $1.307615357$ & $F_6^{\phi}/F_0^{\Tilde{\phi}}$ & $-0.04576648210$ & $-0.9964527481$ \\ \hline
        $F_7^{\Tilde{\phi}}/F_0^{\Tilde{\phi}}$ & $-0.01265605076$ & $-0.3804076087$ & \multicolumn{3}{c}{}\\ \cline{1-3}
    \end{tabular}
    \caption{One-particle form factors of the $\sigma, \sigma'$ and the $\mu, \mu'$ fields in the disordered phase.}
    \label{tab:1partFFs}
\end{table}

For the two-particle form factors, we adopt the parameterisation in Ref.~\cite{2022ScPP...12..162C} valid for both local and semi-local fields
\begin{align}
    F_{ab}^\phi(\vartheta) &= F_0^{\Tilde{\phi}} \left(\cosh\frac{\vartheta}{2}\right)^{\eta_{ab}}\frac{F_{ab}^\text{min}(\vartheta)}{D_{ab}(\vartheta)}\sum_{k=0}^{N^\phi_{ab}} a_{ab}^k\cosh^k(\vartheta)\,,
\end{align}
where $\eta_{ab}$ is
\begin{align}
    \eta_{ab} = \left\{\begin{matrix}
        1 & \text{if both $a$ and $b$ are odd and $a \neq b$}\,,\\
        -1 & \text{if both $a$ and $b$ are odd and $a = b$}\,,\\
        0 & \text{otherwise}\,,
    \end{matrix}\right.
\end{align}
and $N_{ab}^\phi$ can be given as
\begin{align}
    N^\phi_{ab} = \left\lfloor C_{ab}^{-1} - \frac{\delta_{ab}}{2} \right\rfloor\,.
\end{align}
The $a_{ab}^k$ coefficients are summarized in Table \ref{tab:2partFFCoeffs} up to $F_{33}^{\Tilde{\phi}}$ and $F_{34}^\phi$, and are in perfect agreement with Ref.~\cite{2022ScPP...12..162C}.

\begin{table}[h!]
    \centering
    \renewcommand*{\arraystretch}{1.2}
    \begin{tabular}{||c|c|c||c|c|c||} \hline
         $a_{ab}^k$ & $\mu$ & $\mu'$ & $a_{ab}^k$ & $\sigma$ & $\sigma'$ \\ \hline \hline
         $a_{11}^0$ & $1.420276626$ & $13.30740265$ & $a_{12}^0$ & $4.978541090$ & $52.25773665$ \\ \hline
         $a_{11}^1$ & $0.4202766255$ & $12.30740265$ & $a_{12}^1$ & $0.9982554599$ & $44.78750084$ \\ \hline
         $a_{22}^0$ & $4.559468390$ & $56.54780749$ & $a_{14}^0$ & $200.4474457$ & $3123.315720$ \\ \hline
         $a_{22}^1$ & $0.6202373781$ & $42.63409340$ & $a_{14}^1$ & $223.0108845$ & $5080.113042$ \\ \hline
         $a_{13}^0$ & $14.84366839$ & $224.2826676$ & $a_{14}^2$ & $26.63818946$ & $1975.217039$ \\ \hline
         $a_{13}^1$ & $2.185830626$ & $150.2503886$ & $a_{15}^0$ & $-15.21895814$ & $-308.5942414$ \\ \hline
         $a_{24}^0$ & $23.43194594$ & $402.4891521$ & $a_{15}^1$ & $-15.83835748$ & $-454.9024585$ \\ \hline
         $a_{24}^1$ & $22.50682406$ & $609.3604342$ & $a_{15}^2$ & $-1.514766852$ & $-154.0030603$ \\ \hline
         $a_{24}^2$ & $1.871095923$ & $212.5642059$ & $a_{23}^0$ & $40.01763859$ & $646.8519190$ \\ \hline
         $a_{33}^0$ & $54.19198596$ & $1079.645610$ & $a_{23}^1$ & $38.18438631$ & $997.8394045$ \\ \hline
         $a_{33}^1$ & $89.32433223$ & $2444.863959$ & $a_{23}^2$ & $3.402096435$ & $358.2858687$ \\ \hline
         $a_{33}^2$ & $38.47737358$ & $1744.586809$ & $a_{34}^0$ & $232.8536157$ & $4728.486495$ \\ \hline
         $a_{33}^3$ & $2.345027311$ & $378.3684595$ & $a_{34}^1$ & $371.7279303$ & $10525.71268$ \\ \hline
         \multicolumn{3}{c|}{} & $a_{34}^2$ & $156.0832308$ & $7347.672515$ \\ \cline{4-6}
         \multicolumn{3}{c|}{} & $a_{34}^3$ & $8.876050464$ & $1544.890872$ \\ \cline{4-6}
    \end{tabular}
    \caption{Two-particle form factor coefficients of the $\sigma, \sigma'$ and the $\mu, \mu'$ fields in the disordered phase.}
    \label{tab:2partFFCoeffs}
\end{table}

\subsection{Construction of general multi-particle form factors}

From the $A_1$ tower, every multi-particle form factor can be constructed using the bound state singularity equation. However, evaluating the necessary residue terms requires substantial analytic effort, which can be made significantly more effective using a numerical procedure outlined below. As shown before, determining a form factor involves calculating a finite number of coefficients, $C_{\{\lambda\}}$ of the $Q^\phi$ polynomials. A generic form factor polynomial $Q_{\textbf{1}^{n_1},\ldots,\textbf{7}^{n_7}}^\phi$ can be parameterised as
\begin{align}
    \begin{aligned}
        Q_{\textbf{1}^{n_1},\ldots,\textbf{7}^{n_7}}^\phi (x^{(1)}_1,\ldots,x^{(1)}_{n_1},\ldots,x^{(7)}_1,\ldots,x^{(7)}_{n_7}) = \sum_{\{\lambda\}}C_{\{\lambda\}}\prod_{a = 1}^7 \sigma^{(a)}_{\lambda_a}(x^{(a)}_1,\ldots,x^{(a)}_{n_a})\,,
    \end{aligned}
\end{align}
where the product runs over the particle species, $a$, and each $\lambda_a$ is an integer partition. The bound state singularity equations representing the fusion
\begin{align}
    F_{1,\ldots,1}^\phi\rightarrow F_{a_1,\ldots,a_n}^\phi
\end{align}
can be evaluated numerically. Separating the $D$ functions containing the bound state singularities, the rest of the form factor is regular at the fusing angles and, therefore, can be evaluated by an explicit numerical substitution at any $(\vartheta_1,\ldots,\vartheta_n)$ rapidity settings, resulting in a linear system of equations in terms of the $C_{\{\lambda\}}$ variables. The idea is to choose several random settings for $(\vartheta_1,\ldots,\vartheta_n)$ rapidity settings, which are at least as numerous as the independent $C_{\{\lambda\}}$ coefficients in the target form factor and solve the (possibly over-determined) linear system using e.g. QR-decomposition. 

For this procedure, it is necessary to set up a suitable parameterisation of the general multi-particle form factor. The Ansatz is an appropriate generalisation of the previously studied $A_1$ and $A_2$ towers:
\begin{align}
    \begin{aligned}
        F_{a_1,\ldots,a_n}^\phi(\vartheta_1,\ldots,\vartheta_n) &= H_{a_1,\ldots,a_n}^\phi \frac{Q_{a_1,\ldots,a_n}^\phi(x_1,\ldots,x_n)}{\omega_{a_1,\ldots,a_n}^\phi(x_1,\ldots,x_n)}\prod_{k<l}^n\frac{F_{a_ka_l}^\text{min}(\vartheta_k - \vartheta_l)}{D_{a_ka_l}(\vartheta_k - \vartheta_l)(x_k+x_l)^{\delta_{a_ka_l}}}\,,
    \end{aligned}
\end{align}
where $H_{a_1,\ldots,a_n}^\phi$ is some normalisation factor chosen by convention, $Q_{a_1,\ldots,a_n}^\phi(x_1,\ldots,x_n)$ is a homogeneous polynomial symmetric in the variables corresponding to the same particle species, and $\omega_{a_1,\ldots,a_n}^\phi(x_1,\ldots,x_n)$ is some appropriately chosen monomial of the $x_k$'s, which can be fixed by analysing the general kinematic and bound state singularity equations.

We remark that one can generalise our previous choice of the coefficients $H$ to
\begin{align}
    H_{a_1,\ldots,a_n}^\phi = F_0^\phi \prod_{k<l}^n \kappa_{a_ka_l}^{-1/2}\prod_{k=1}^n \kappa_{a_ka_k}^{1/4}\,,
\end{align}
which is most suitable for analytic calculations; however, for the numerical approach, it can be fixed simply as $F_0^\phi$. The minimal choice for the monomial
\begin{align}
    \omega_{a_1,\ldots,a_n}^\phi(x_1,\ldots,x_n) = x_1^{\deg_{x_1}(\omega)}\cdot\ldots\cdot x_n^{\deg_{x_n}(\omega)}
\end{align}
can fixed in terms its degrees $\deg_{x_n}$. For the $\Theta$ field, these are given by
\begin{align}
    \begin{aligned}
        \deg_{x_k}\left(\omega_{a_1,\ldots,a_n}^\Theta(x_1,\ldots,x_n)\right) = \left\lfloor \sum_{l=1}^n\left(C^{-1}_{a_ka_l}-\delta_{a_ka_l}\right) - \left(C^{-1}_{a_ka_k}-1\right)\right\rfloor\,,
    \end{aligned}
\end{align}
while for the order and disorder fields,
\begin{align}
    \begin{aligned}
        \deg_{x_k}\left(\omega_{a_1,\ldots,a_n}^\phi(x_1,\ldots,x_n)\right) = \sum_{l=1}^n\left(C^{-1}_{a_ka_l}-\delta_{a_ka_l}\right) - \left(C^{-1}_{a_ka_k}-1\right)\,.
    \end{aligned}
\end{align}

The partial degrees of the $Q_{a_1,\ldots,a_n}^\phi$ polynomials can be derived from Lorentz invariance and the convergence of the limit in the clustering equation. For the $\Theta$ field one obtains
\begin{align}
    \begin{aligned}
        \deg_{x_1,\ldots,x_n}&\left(Q_{a_1,\ldots,a_n,b_1,\ldots,b_m}^\Theta(x_1,\ldots,x_n,y_1,\ldots,y_m)\right) =\\ &\sum_{k<l}^n\left(2C^{-1}_{a_ka_l}-\delta_{a_ka_l}\right) + \sum_{k=1}^n\sum_{l=1}^m\left(2C^{-1}_{a_kb_l}-\delta_{a_kb_l}\right) - \left\lfloor \frac{1}{2}\sum_{k=1}^n q_{a_k}^{(9)}\right\rfloor\,,
    \end{aligned}
\end{align}
where
\begin{align}
    q_a^{(9)} = \left\{\begin{matrix}
        0 & \text{if $a$ is even}\,,\\
        1 & \text{if $a$ is odd}\,,
    \end{matrix}\right.
\end{align}
or equivalently,
\begin{align}
    \begin{aligned}
        \deg_{n_1',\ldots,n_7'}\left(Q_{\textbf{1}^{n_1},\ldots,\textbf{7}^{n_7}}^\Theta\right) =& \sum_{a=1}^7\left(2C^{-1}_{aa}-1\right)n_a'\left(n_a-\frac{n_a'}{2}-\frac{1}{2}\right)\\
        &+ \sum_{a<b}^7 2C^{-1}_{ab} \left(n_a'n_b+n_an_b'-n_a'n_b'\right) - \left\lfloor \frac{1}{2}\sum_{a\,\text{odd}}^n n_a'\right\rfloor
    \end{aligned}
\end{align}
if the $\{a_1,\ldots,a_n,b_1,\ldots,b_m\}$ multi-particle state is even (otherwise the form factor vanishes). 

For the $\sigma$, $\sigma'$, $\mu$ and $\mu'$ fields, the partial degrees are
\begin{align}
    \begin{aligned}
        \deg_{x_1,\ldots,x_n}&\left(Q_{a_1,\ldots,a_n,b_1,\ldots,b_m}^\phi(x_1,\ldots,x_n,y_1,\ldots,y_m)\right) =\\ &\sum_{k<l}^n\left(2C^{-1}_{a_ka_l}-\delta_{a_ka_l}\right) + \sum_{k=1}^n\sum_{l=1}^m\left(2C^{-1}_{a_kb_l}-\delta_{a_kb_l}\right)\,,
    \end{aligned}
\end{align}
or equivalently,
\begin{align}
    \begin{aligned}
        \deg_{n_1',\ldots,n_7'}\left(Q_{\textbf{1}^{n_1},\ldots,\textbf{7}^{n_7}}^\phi\right) =& \sum_{a=1}^7\left(2C^{-1}_{aa}-1\right)n_a'\left(n_a-\frac{n_a'}{2}-\frac{1}{2}\right)\\
        &+ \sum_{a<b}^7 2C^{-1}_{ab} \left(n_a'n_b+n_an_b'-n_a'n_b'\right)\,.
    \end{aligned}
\end{align}
Furthermore, the form factors of the $\Theta$ field are proportional to $P^2$, which yields the following factorisation
\begin{align}
    \begin{aligned}
    &Q_{\textbf{1}^{n_1},\ldots,\textbf{7}^{n_7}}^\Theta =\\ &\left\{\sum_{a=1}^7m_a^2\sigma_1^{(a)}\sigma_{n_a-1}^{(a)}\prod_{b\neq a}^7\sigma_{n_b}^{(b)}+\sum_{a<b}^7m_am_b\left(\sigma_1^{(a)}\sigma_{n_a}^{(a)}\sigma_{n_b-1}^{(b)} + \sigma_1^{(b)}\sigma_{n_b}^{(b)}\sigma_{n_a-1}^{(a)}\right)\prod_{c\neq a,b}^7\sigma_{n_c}^{(c)}\right\}\hat{Q}_{\textbf{1}^{n_1},\ldots,\textbf{7}^{n_7}}^\Theta
    \end{aligned}
\end{align}
further reducing the number of independent coefficients.


\section{Conclusions and outlook}\label{sec:discussion}

The main result of the present work is a systematic approach to the construction of all form factors of the trace of the stress-energy tensor $\Theta$ and the order/disorder fields $\sigma,\sigma'/\mu,\mu'$ in the thermal perturbation of the tricritical Ising model, which corresponds to the $E_7$ factorised scattering theory. This required determining the form factors containing only the fundamental excitation $A_1$ a.k.a.~the $A_1$ tower since all other form factors can be obtained using the form factor bound state singularity equation and the bootstrap structure of the $E_7$ $S$ matrix. We started by determining the $A_2$ tower form factors containing only the second particle $A_2$. This task is made simpler by the self-fusion pole of $A_2$. In conjunction with the bound on the asymptotic growth of the form factors and the clustering property, this allowed fixing all form factors in terms of the single particle $A_2$ form factor $F_2^\phi$ and the vacuum expectation value $F_0^\phi$. We then presented a simple procedure to determine the ratio $F_2^\phi/F_0^\phi$ ratio, which involved finding a minimal solvable subsystem of form factors involving the $A_1A_1$ form factor, which can also be helpful for form factors of other fields of the $E_7$ model. Together with the exact vacuum expectation values available from \cite{russian2}, this fixes all the form factors which involve only even particles.

To determine the $A_1$ tower, finding a minimal set of recursive equations containing all the information encoded in the rich bootstrap structure of the $E_7$ scattering theory was necessary. We found that such a system can be obtained by combining the kinematic singularity equation with a further relation resulting from the bound state singularity equations resulting from equating two different orderings of combining an $A_1A_1\rightarrow A_2$ and an $A_1A_2\rightarrow A_3$ fusion, which we called the three-$A_1$ fusion equation. To our knowledge, this type of recursive structure has not been found before. We gave a mathematical proof that these equations were enough to fully determine all even elements of the $A_1$ tower from the $6$-particle level upwards, while the  $4$-particle form factor can be determined using the $A_2$ tower determined earlier. Once this is accomplished, the clustering property determines odd elements of the $A_1$ tower.

We cross-checked our results by computing all one-particle and several two-particle form factors involving higher excitations in the $E_7$ spectrum, which agreed perfectly with previous results \cite{AMV, 2022ScPP...12..162C}.

An interesting open problem is to find a systematic construction for the form factor of the vacancy density $t$, which is the only relevant field whose form factors are presently unknown. The challenge is that it has a higher scaling dimension than $\Theta$ and the order/disorder fields, which implies that the asymptotic growth of its form factors is less constrained. 

The results of the present work can be used to compute correlation functions and dynamical structure factors in the thermal perturbation of the tricritical Ising model (cf. \cite{2022ScPP...12..162C,2023JPhA...56L3001L}). Additionally, the novel recursive structure found in our work can help to find a systematic construction of form factors in other exactly integrable models.

\paragraph{Acknowledgements}

The authors are grateful to M.~Lencs\'es for valuable discussions. This work was supported by the  National Research, Development and Innovation Office of Hungary (NKFIH) through the OTKA Grant K 138606. GT was also partially supported by the NKFIH grant ``Quantum Information National Laboratory'' (Grant No. 2022-2.1.1-NL-2022-00004), while BF was partially supported by the ÚNKP-23-2-III-BME-132 New National Excellence Program of the Ministry for Culture and Innovation from the National Research, Development and Innovation Fund. 

\clearpage

\appendix

\section{The $E_7$ $S$-matrix and form factor building blocks}\label{sec:building_blocks}

\subsection{The exact $S$-matrix}

The two-particle amplitudes in a diagonal scattering theory of self-conjugate particles can be written as
\begin{align}
    \begin{aligned}
    S_{ab}(\vartheta) &= (-1)^{\delta_{ab}} \prod_{\alpha \in\mathcal{A}_{ab}} (-f_\alpha(\vartheta))^{p_{ab}(\alpha)}\\
    &= \prod_{\alpha \in \overline{\mathcal{A}_{ab}}} (-f_\alpha(\vartheta))^{p_{ab}(\alpha)}\,,
    \end{aligned}
\end{align}
where $\mathcal{A}_{ab}$ is the collection of the poles and $p_{ab}(\alpha)$ is the multiplicity of the pole at $\alpha i \pi$, and we introduced the
\begin{align}
    \overline{\mathcal{A}_{ab}} &= \left\{\begin{matrix}\mathcal{A}_{ab}\cup\{0\}\,,\quad \text{if}\quad{a = b}\,,\\
    \mathcal{A}_{ab}\,,\quad \text{otherwise}\end{matrix}\right.
\end{align}
notation for convenience. The $S$-matrix building blocks can be given as
\begin{align}
    \begin{aligned}
        f_\alpha(\vartheta) &= \frac{\tanh\frac{1}{2}(\vartheta + i\pi \alpha)}{\tanh\frac{1}{2}(\vartheta - i\pi \alpha)}\\
        &= -\exp\left\{2\int_0^\infty \frac{dt}{t}\frac{\cosh\left(\alpha - \frac{1}{2}\right)t}{\cosh\frac{t}{2}}\sinh\frac{\vartheta t}{i\pi}\right\}\,.
    \end{aligned}
\end{align}
For the $E_7$ model, the $S_{ab}$ are specified in Table \ref{tab:E7Aab}, where we use the notation
\begin{align}
    \stackrel{\textbf{a}}{(\xi)^{p}} \equiv \left(-f_{\xi/18}(\vartheta)\right)^{p}
\end{align}
for a block corresponding to a pole at $\alpha=\xi/18$ with multiplicity $p$, which gives rise to the bound state $A_a$ (for poles which do not correspond to bound states, the top index $\textbf{a}$ is absent). The absolute values of the $3$-point couplings can be computed from the residue of the bound state poles of the $S$-matrix:
\begin{align} \label{eq:resGamma}
    |\Gamma_{ab}^c|^2 &= -i\lim_{\vartheta-iu_{ab}^c}(\vartheta-iu_{ab}^c)S_{ab}(\vartheta)\,.
\end{align}
Regarding their phases, a consistent setting is given by choosing every $\Gamma_{ab}^c$ real and positive.
\begin{table}[t]
    \centering
    \renewcommand*{\arraystretch}{2.0}
    \begin{tabular}{|c|l|} \hline
         $a\quad b$ & \multicolumn{1}{c|}{$S_{ab}$} \\ \hline \hline
         1\quad 1 & $-\stackrel{\textbf{2}}{(10)}\stackrel{\textbf{4}}{(2)}$ \\ \hline
         1\quad 2 & $\stackrel{\textbf{1}}{(13)}\stackrel{\textbf{3}}{(7)}$ \\ \hline
         1\quad 3 & $\stackrel{\textbf{2}}{(14)}\stackrel{\textbf{4}}{(10)}\stackrel{\textbf{5}}{(6)}$ \\ \hline
         1\quad 4 & $\stackrel{\textbf{1}}{(17)}\stackrel{\textbf{3}}{(11)}\stackrel{\textbf{6}}{(3)}(9)$ \\ \hline
         1\quad 5 & $\stackrel{\textbf{3}}{(14)}\stackrel{\textbf{6}}{(8)}(6)^2$ \\ \hline
         1\quad 6 & $\stackrel{\textbf{4}}{(16)}\stackrel{\textbf{5}}{(12)}\stackrel{\textbf{7}}{(4)}(10)^2$ \\ \hline
         1\quad 7 & $\stackrel{\textbf{6}}{(15)}(5)^2(7)^2(9)$ \\ \hline
         2\quad 2 & $-\stackrel{\textbf{2}}{(12)}\stackrel{\textbf{4}}{(8)}\stackrel{\textbf{5}}{(2)}$ \\ \hline
         2\quad 3 & $\stackrel{\textbf{1}}{(15)}\stackrel{\textbf{3}}{(11)}\stackrel{\textbf{6}}{(5)}(9)$ \\ \hline
         2\quad 4 & $\stackrel{\textbf{2}}{(14)}\stackrel{\textbf{5}}{(8)}(6)^2$ \\ \hline
         2\quad 5 & $\stackrel{\textbf{2}}{(17)}\stackrel{\textbf{4}}{(13)}\stackrel{\textbf{7}}{(3)}(7)^2(9)$ \\ \hline
         2\quad 6 & $\stackrel{\textbf{3}}{(15)}(7)^2(5)^2(9)$ \\ \hline
         2\quad 7 & $\stackrel{\textbf{5}}{(16)}\stackrel{\textbf{7}}{(10)^3}(4)^2(6)^2$ \\ \hline
         3\quad 3 & $-\stackrel{\textbf{2}}{(14)}\stackrel{\textbf{7}}{(2)}(8)^2(12)^2$ \\ \hline
    \end{tabular}
    ~~~~~~~~~~~~
    \begin{tabular}{|c|l|} \hline
         $a\quad b$ & \multicolumn{1}{c|}{$S_{ab}$}\\ \hline \hline
         3\quad 4 & $\stackrel{\textbf{1}}{(15)}(5)^2(7)^2(9)$ \\ \hline
         3\quad 5 & $\stackrel{\textbf{1}}{(16)}\stackrel{\textbf{6}}{(10)^3}(4)^2(6)^2$ \\ \hline
         3\quad 6 & $\stackrel{\textbf{2}}{(16)}\stackrel{\textbf{5}}{(12)^3}\stackrel{\textbf{7}}{(8)^3}(4)^2$ \\ \hline
         3\quad 7 & $\stackrel{\textbf{3}}{(17)}\stackrel{\textbf{6}}{(13)^3}(3)^2(7)^4(9)^2$ \\ \hline
         4\quad 4 & $-\stackrel{\textbf{4}}{(12)}\stackrel{\textbf{5}}{(10)^3}\stackrel{\textbf{7}}{(4)}(2)^2$ \\ \hline
         4\quad 5 & $\stackrel{\textbf{2}}{(15)}\stackrel{\textbf{4}}{(13)^3}\stackrel{\textbf{7}}{(7)^3}(9)$ \\ \hline
         4\quad 6 & $\stackrel{\textbf{1}}{(17)}\stackrel{\textbf{6}}{(11)^3}(3)^2(5)^2(9)^2$ \\ \hline
         4\quad 7 & $\stackrel{\textbf{4}}{(16)}\stackrel{\textbf{5}}{(14)^3}(6)^4(8)^4$ \\ \hline
         5\quad 5 & $-\stackrel{\textbf{5}}{(12)^3}(2)^2(4)^2(8)^4$ \\ \hline
         5\quad 6 & $\stackrel{\textbf{1}}{(16)}\stackrel{\textbf{3}}{(14)^3}(6)^4(8)^4$ \\ \hline
         5\quad 7 & $\stackrel{\textbf{2}}{(17)}\stackrel{\textbf{4}}{(15)^3}\stackrel{\textbf{7}}{(11)^5}(5)^4(9)^3$ \\ \hline
         6\quad 6 & $-\stackrel{\textbf{4}}{(14)^3}\stackrel{\textbf{7}}{(10)^5}(12)^4(16)^2$ \\ \hline
         6\quad 7 & $\stackrel{\textbf{1}}{(17)}\stackrel{\textbf{3}}{(15)^3}\stackrel{\textbf{6}}{(13)^5}(5)^6(9)^3$ \\ \hline
         7\quad 7 & $-\stackrel{\textbf{2}}{(16)^3}\stackrel{\textbf{5}}{(14)^5}\stackrel{\textbf{7}}{(12)^7}(8)^8$ \\ \hline
    \end{tabular}
    \caption{The two-particle $S$-matrix amplitudes in the $E_7$ scattering theory.}
    \label{tab:E7Aab}
\end{table}

\subsection{Form factor building blocks}

Using the notations introduced above, the $D_{ab}(\vartheta)$ polynomials encoding the bound state singularity structure are defined as
\begin{align}
    D_{ab}(\vartheta) &= \prod_{\alpha\in\mathcal{A}_{ab}}(P_\alpha(\vartheta))^{i_{ab}(\alpha)}(P_{1-\alpha}(\vartheta))^{j_{ab}(\alpha)}\,,
\end{align}
where
\begin{align}
    P_\alpha &= \frac{\cos(\pi\alpha) - \cosh(\vartheta)}{2\cos^2\left(\frac{\pi\alpha}{2}\right)}\,,\\
    i_{ab}(\alpha) &= n+1\,,\quad j_{ab}(\alpha) = n\,,\quad \text{if}\quad p_{ab}(\alpha) = 2n+1\,,\\
    i_{ab}(\alpha) &= n\,,\quad j_{ab}(\alpha) = n\,,\quad \text{if}\quad p_{ab}(\alpha) = 2n\,.
\end{align}
The minimal form factor can be written as
\begin{align}
    \begin{aligned}
        F_{ab}^\text{min}(\vartheta) &= \prod_{\alpha\in\overline{\mathcal{A}_{ab}}} (g_\alpha(\vartheta))^{p_{ab}(\alpha)}\\
        &= \left(-i\sinh\left(\frac{\vartheta}{2}\right)\right)^{\delta_{ab}} \prod_{\alpha\in\mathcal{A}_{ab}} (g_\alpha(\vartheta))^{p_{ab}(\alpha)}\,,
    \label{eq:minFab_soln}\end{aligned}
\end{align}
where
\begin{align}
    g_\alpha(\vartheta) &= \exp\left\{2\int_0^\infty\frac{dt}{t}\frac{\cosh\left(\alpha-\frac{1}{2}\right)t}{\cosh\frac{t}{2}\sinh t}\sin^2\frac{(i\pi - \vartheta)t}{2\pi}\right\}\,. \label{Eq:galpha}
\end{align}
To improve the numerical convergence of the integral representation of $g_\alpha(\vartheta)$, it is best to use the identity
\begin{align}
    \begin{aligned}
        g_\alpha(\vartheta) &= \prod_{k=0}^{N-1} \left[\frac{\left[1+\left(\frac{\Hat{\vartheta}/2\pi}{k + 1 - \alpha/2}\right)^2\right]\left[1+\left(\frac{\Hat{\vartheta}/2\pi}{k + 1/2 + \alpha/2}\right)^2\right]}{\left[1+\left(\frac{\Hat{\vartheta}/2\pi}{k + 1 + \alpha/2}\right)^2\right]\left[1+\left(\frac{\Hat{\vartheta}/2\pi}{k + 3/2 - \alpha/2}\right)^2\right]}\right]^{k+1}\\
        &\quad\times\exp\left\{2\int_0^\infty\frac{dt}{t}\frac{\cosh\left(\alpha-\frac{1}{2}\right)t}{\cosh\frac{t}{2}\sinh t}(N+1-Ne^{-2t})e^{-2Nt}\sin^2\frac{(i\pi - \vartheta)t}{2\pi}\right\}
    \end{aligned}
\end{align}
with some conveniently chosen $N$ (for numerical evaluations we used $N = 40$). The normalisation of both $F_{ab}^\text{min}(\vartheta)$ and $D_{ab}(\vartheta)$ are chosen so that
\begin{align}
    D_{ab}(i\pi) = 1\quad\text{and}\quad F_{ab}^\text{min}(i\pi) = 1\,.
\end{align}
The functions $g_\alpha(\vartheta)$ satisfy
\begin{align} \label{Eq:gblockrel}
        g_\alpha(\vartheta) &= -f_\alpha(\vartheta)g_\alpha(-\vartheta)\,,\\
        g_\alpha(i\pi+\vartheta) &= g_\alpha(i\pi-\vartheta)\,,
\end{align}
which guarantees that \eqref{eq:minFab_soln} solves the defining relations of the minimal form factors \eqref{Eq:minFFSrel}, and are free of poles and zeros in the strip $\mathcal{S} = \{ \text{Im }\vartheta \in (0,\pi) \}$. 

Moreover, the $g_\alpha(\vartheta)$ building blocks satisfy some further identities that were used in our calculations:
\begin{align}
    g_\alpha(\vartheta) &= g_{1-\alpha}(\vartheta)\,,\\
    g_0(\vartheta) &= -i\sinh\frac{\vartheta}{2}\,,\\
    g_\alpha(\vartheta+i\pi\kappa)g_\alpha(\vartheta-i\pi\kappa) &= \frac{g_\alpha(i\pi\kappa)g_\alpha(-i\pi\kappa)}{g_{\alpha+\kappa}(0)g_{\alpha-\kappa}(0)}g_{\alpha+\kappa}(\vartheta)g_{\alpha-\kappa}(\vartheta)\,,\\
    g_{\alpha}(\vartheta)g_{-\alpha}(\vartheta) &= -\frac{\sinh\frac{1}{2}(\vartheta-i\alpha\pi)\sinh\frac{1}{2}(\vartheta+i\alpha\pi)}{\cos^2\frac{\pi\alpha}{2}}\,.
\end{align}
Their asymptotic behaviour for $|\vartheta|\rightarrow \infty$ is
\begin{align}
    g_\alpha(\vartheta) &\sim \mathcal{N}_\alpha\exp\left(\frac{|\vartheta|}{2} - \frac{i\pi}{2}\right)\,,
\end{align}
where
\begin{align}
    \begin{aligned}
        \mathcal{N}_\alpha &= \exp\left\{\int_0^\infty\frac{dt}{t}\left[\frac{\cosh\left(\alpha-\frac{1}{2}\right)t}{\cosh\frac{t}{2}\sinh t} - \frac{1}{t}\right]\right\}\,.
    \end{aligned}
\end{align}


\section{Proof of completeness}\label{sec:proof}

In this appendix, we prove that that the kinematic equation along with the $A_1^3\rightarrow A_1^3$ bound state singularity equation (eqs.~\eqref{Eq:KPTheta} and \eqref{Eq:BSPTheta} for the $\Theta$, or eqs.~\eqref{Eq:KPSL} and \eqref{Eq:BSPSL}  or the semi-local fields) completely determines the polynomials $Q_{\textbf{1}^{2n}}^\phi$ for $2n \geq 6$.

\begin{figure}[t]
    \centering
    \includegraphics{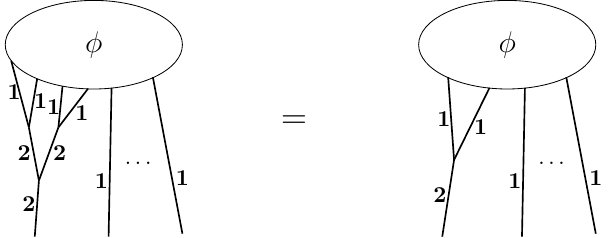}

    \vspace{0.3cm}
    \includegraphics{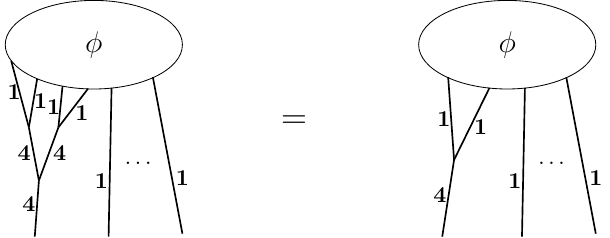}

    \caption{Schematic illustration of fusion chains \eqref{Eq:FCC}.}
    \label{fig:1111to11}
\end{figure}

First, we remark that none of the $A_1^n\rightarrow A_1^m$ four-fold fusion equations gives further information beyond the kinematic singularity equation and the $A_1^3\rightarrow A_1^3$ fusion. For example, the fusing angles corresponding to the
\begin{align}
    \begin{aligned} \label{Eq:FCC}
        &(A_1\times A_1)\times(A_1\times A_1)\rightarrow A_2\times A_2\rightarrow A_2 \leftarrow A_1\times A_1\,,\quad \text{and}\\
        &(A_1\times A_1)\times(A_1\times A_1)\rightarrow A_4\times A_4\rightarrow A_4 \leftarrow A_1\times A_1
    \end{aligned}
\end{align}
fusion chains (illustrated in Fig.~\ref{fig:1111to11}) can be obtained by applying crossing symmetry to one of the particles in the $A_1^3 \rightarrow A_1^3$ fusion. Therefore, consecutive application of the $A_1^3\rightarrow A_1^3$ fusion and the kinematic singularity equation at some special angles leads to the same equation as the above fusion chains. Similarly, it can be established that none of the $A_1^5\rightarrow A_1$ fusion chains gives further information either. We also verified some of the more complex $A_1^n\rightarrow A_1^m$ fusion chains. This leads to the conjecture that the kinematic singularity equation and the $A_1^3\rightarrow A_1^3$ bound state singularity equation capture all the nontrivial constraints the $E_7$ bootstrap poses.

This conjecture can be made precise as the following theorem:
\begin{theorem}
    For $2n > 6$, there is no non-zero kernel of the kinematic singularity equation that satisfies the $A_1^3\rightarrow A_1^3$ fusion equation and respects the partial degree restrictions in eqs.~\eqref{Eq:partDegTh} or \eqref{Eq:partDegSl}.
\end{theorem}
This theorem is equivalent to the statement that beyond the six-particle level, the kinematic singularity equation and the $A_1^3\rightarrow A_1^3$ fusion uniquely determine all members of the $ A_1$ tower.

\begin{proof}
    Let $\mathcal{K}_{2n}$ be the a polynomial in the aforementioned kernel, so
    \begin{align}\label{Eq:AppKinKernel}
        \mathcal{K}_{2n}(-x,x,x_3,\ldots,x_{2n}) &= 0
    \end{align}
    and
    \begin{align}
        \begin{aligned} \label{Eq:AppKA13toA13}
            &e^{\frac{8}{18}i\pi(n-1)}\prod_{k=4}^{2n}[12]_k[-16]_k\, \mathcal{K}_{2n}\left(e^{\frac{4}{18}i\pi}x, e^{\frac{2}{18}i\pi}x, e^{-\frac{8}{18}i\pi}x, x_4,\ldots,x_{2n}\right)=\\
            &= e^{-\frac{8}{18}i\pi(n-1)}\prod_{k=4}^{2n}[-12]_k[16]_k\, \mathcal{K}_{2n}\left(e^{\frac{8}{18}i\pi}x, e^{-\frac{2}{18}i\pi}x, e^{-\frac{4}{18}i\pi}x, x_4,\ldots,x_{2n}\right)\,.
        \end{aligned}
    \end{align}
    In the course of this proof, we denote for simplicity 
    \begin{align}
        [\alpha]_k = x - e^{\alpha i\pi/18} x_k\,.
    \end{align}
    Eqn. \eqref{Eq:AppKinKernel} implies that $\mathcal{K}_{2n}$ can be written as
    \begin{align}
        \mathcal{K}_{2n}(x_1,\ldots,x_{2n}) &= L_{2n}(x_1,\ldots,x_{2n})\prod_{k<l}^{2n}(x_k+x_l)\,,
    \end{align}
    where $L_{2n}$ is some symmetric homogeneous polynomial.
    Comparing to eq.~\eqref{Eq:partDegSl}, we get the following partial degree restriction for the $L_{2n}$ polynomials
    \begin{align} \label{Eq:partDegL}
        \deg_k\left(L_{2n}\right) \leq k\left(2n - \frac{k}{2}-\frac{1}{2}\right)\,.
    \end{align}
    Substituting back into eq.~\eqref{Eq:AppKA13toA13} results in the condition 
    \begin{align}
        \begin{aligned}
            &e^{\frac{4}{18}i\pi(n-1)}\prod_{k=4}^{2n}[14]_k[12]_k[-10]_k\, L_{2n}\left(e^{\frac{4}{18}i\pi}x, e^{\frac{2}{18}i\pi}x, e^{-\frac{8}{18}i\pi}x, x_4,\ldots,x_{2n}\right)=\\
            &= e^{-\frac{4}{18}i\pi(n-1)}\prod_{k=4}^{2n}[-14]_k[-12]_k[10]_k\, L_{2n}\left(e^{\frac{8}{18}i\pi}x, e^{-\frac{2}{18}i\pi}x, e^{-\frac{4}{18}i\pi}x, x_4,\ldots,x_{2n}\right)
        \end{aligned}
    \end{align}
    for $L_{2n}$. Both sides are polynomials in $x$, which must have the same roots, implying 
    \begin{align} \label{Eq:AppKerLlhs}
        L_{2n}\left(e^{\frac{4}{18}i\pi}x, e^{\frac{2}{18}i\pi}x, e^{-\frac{8}{18}i\pi}x, x_4,\ldots,x_{2n}\right) &\sim \prod_{k=4}^{2n}[-14]_k[-12]_k[10]_k\,,
    \end{align}
    and
    \begin{align} \label{Eq:AppKerLrhs}
        L_{2n}\left(e^{\frac{8}{18}i\pi}x, e^{-\frac{2}{18}i\pi}x, e^{-\frac{4}{18}i\pi}x, x_4,\ldots,x_{2n}\right) &\sim \prod_{k=4}^{2n}[14]_k[12]_k[-10]_k\,,
    \end{align}
    where $L\sim Q$ means that there exist a polynomial $P$ such that $L = P\cdot Q$. Substituting particular arguments and using the symmetric property of $L_{2n}$ we obtain the conditions
    \begin{align}
        \begin{aligned}
            &L_{2n}\left(e^{\frac{6}{18}i\pi}x, e^{\frac{4}{18}i\pi}x, e^{-\frac{4}{18}i\pi}x, e^{-\frac{6}{18}i\pi}x, x_5,\ldots, x_{2n}\right)\\
            &\quad \sim \left(e^{\frac{2}{18}i\pi}x - e^{-\frac{14}{18}i\pi}e^{-\frac{4}{18}i\pi}x\right)\left(e^{\frac{2}{18}i\pi}x - e^{-\frac{12}{18}i\pi}e^{-\frac{4}{18}i\pi}x\right)\left(e^{\frac{2}{18}i\pi}x - e^{\frac{10}{18}i\pi}e^{-\frac{4}{18}i\pi}x\right)\times\\ &\quad\quad\times\prod_{k=5}^{2n}\left(e^{\frac{2}{18}i\pi}x - e^{-\frac{14}{18}i\pi}x_k\right)\left(e^{\frac{2}{18}i\pi}x - e^{-\frac{12}{18}i\pi}x_k\right)\left(e^{\frac{2}{18}i\pi}x - e^{\frac{10}{18}i\pi}x_k\right)\\
            &\quad \sim x^3\prod_{k=5}^{2n}[-16]_k[-14]_k[8]_k
        \end{aligned}
    \end{align}
    and
    \begin{align}
        \begin{aligned}
            &L_{2n}\left(e^{\frac{6}{18}i\pi}x, e^{\frac{4}{18}i\pi}x, e^{-\frac{4}{18}i\pi}x, e^{-\frac{6}{18}i\pi}x, x_5,\ldots, x_{2n}\right)\\
            &\quad \sim \left(e^{-\frac{2}{18}i\pi}x - e^{\frac{14}{18}i\pi}e^{\frac{4}{18}i\pi}x\right)\left(e^{-\frac{2}{18}i\pi}x - e^{\frac{12}{18}i\pi}e^{\frac{4}{18}i\pi}x\right)\left(e^{-\frac{2}{18}i\pi}x - e^{-\frac{10}{18}i\pi}e^{\frac{4}{18}i\pi}x\right)\times\\ &\quad\quad\times\prod_{k=5}^{2n}\left(e^{-\frac{2}{18}i\pi}x - e^{\frac{14}{18}i\pi}x_k\right)\left(e^{-\frac{2}{18}i\pi}x - e^{\frac{12}{18}i\pi}x_k\right)\left(e^{-\frac{2}{18}i\pi}x - e^{-\frac{10}{18}i\pi}x_k\right)\\
            &\quad \sim x^3\prod_{k=5}^{2n}[16]_k[14]_k[-8]_k\,.
        \end{aligned}
    \end{align}
    Consider $L_{2n}$ with the above arguments as a polynomial in $x$. The above equations suggest that it has an (at least) third-order root at $x = 0$ and (at least) first-order roots at
    \begin{align}
        \begin{aligned} \nonumber
            x &= e^{-\frac{16}{18} i\pi}x_k\,,\quad x = e^{-\frac{14}{18} i\pi}x_k\,,\quad x = e^{-\frac{8}{18} i\pi}x_k\,,\\
            x &= e^{\frac{8}{18} i\pi}x_k\,,\quad x = e^{\frac{14}{18} i\pi}x_k\,,\quad\text{and}\quad x = e^{\frac{16}{18} i\pi}x_k\,,
        \end{aligned}
    \end{align}
    therefore, it can be written as
    \begin{align}
        \begin{aligned}\label{eq:L2n}
            &L_{2n}\left(e^{\frac{6}{18}i\pi}x, e^{\frac{4}{18}i\pi}x, e^{-\frac{4}{18}i\pi}x, e^{-\frac{6}{18}i\pi}x, x_5,\ldots, x_{2n}\right)\\
            &\quad \sim x^3\prod_{k=5}^{2n}[16]_k[14]_k[8]_k[-8]_k[-14]_k[-16]_k\,.
        \end{aligned}
    \end{align}
    The further argument is split into three separate cases.
    
    \paragraph{Case 1} The first case assumes that $L_{2n}$ in \eqref{eq:L2n} is not identically $0$. Then, it is a polynomial of $x$ with a degree of at least
    \begin{align}
        3 + 6\cdot(2n - 4) &= 12n - 21\,.
    \end{align}
    Therefore we have a lower bound on the $4$-th order partial degree of a generic $L_{2n}$ polynomial as
    \begin{align} \label{Eq:AppKerLowB}
        \deg_4\left(L_{2n}\right) \geq 12n - 21\,.
    \end{align}
    On the other hand, however, eq.~\eqref{Eq:partDegL} prescribes that
    \begin{align} \label{Eq:AppKerHighB}
        \deg_4\left(L_{2n}\right) \leq 8n - 10\,.
    \end{align}
    Comparing the above two relations, we get
    \begin{align}
        2n \leq \frac{11}{2}\,.
    \end{align}
    
    \paragraph{Case 2} Next, let us consider the case when
    \begin{align} \label{Eq:AppL4eq0}
        L_{2n}\left(e^{\frac{6}{18}i\pi}x, e^{\frac{4}{18}i\pi}x, e^{-\frac{4}{18}i\pi}x, e^{-\frac{6}{18}i\pi}x, x_5,\ldots, x_{2n}\right) = 0\,,
    \end{align}
    but assume that
    \begin{align}
        L_{2n}\left(e^{\frac{4}{18}i\pi}x, e^{\frac{2}{18}i\pi}x, e^{-\frac{8}{18}i\pi}x, x_4,\ldots, x_{2n}\right) \neq 0\,.
    \end{align}
    Then eq.~\eqref{Eq:AppL4eq0} implies that the polynomial $L_{2n}\left(e^{\frac{4}{18}i\pi}x, e^{\frac{2}{18}i\pi}x, e^{-\frac{8}{18}i\pi}x, x_4,\ldots, x_{2n}\right)$ has roots at $x = e^{6i\pi/18}x_k$. Using  eq.~\eqref{Eq:AppKerLlhs} implies
    \begin{align}
        L_{2n}\left(e^{\frac{4}{18}i\pi}x, e^{\frac{2}{18}i\pi}x, e^{-\frac{8}{18}i\pi}x, x_4,\ldots, x_{2n}\right) \sim \prod_{k=4}^{2n} [-14]_k[-12]_k[10]_k[6]_k\,.
    \end{align}
    Due to the assumption that it is not identically $0$, it must be a polynomial of at least order $8n - 12$  in $x$. Therefore, its third partial degree can be estimated as
    \begin{align} \label{Eq:AppKerLowBCase2}
        \deg_3\left(L_{2n}\right) \geq 8n - 12\,.
    \end{align}
    The partial degree restrictions (see eq.~\eqref{Eq:partDegL}) prescribe that
    \begin{align} \label{Eq:AppKerHighBCase2}
        \deg_3\left(L_{2n}\right) \leq 6n - 6\,.
    \end{align}
    Comparing the above equations, we get
    \begin{align}
        2n \leq 6\,.
    \end{align}

    \paragraph{Case 3} The last case is when both
    \begin{align} \label{Eq:AppLZeroLHS}
        L_{2n}\left(e^{\frac{4}{18}i\pi}x, e^{\frac{2}{18}i\pi}x, e^{-\frac{8}{18}i\pi}x, x_4,\ldots, x_{2n}\right) = 0
    \end{align}
    and 
    \begin{align} \label{Eq:AppLZeroRHS}
        L_{2n}\left(e^{\frac{8}{18}i\pi}x, e^{-\frac{2}{18}i\pi}x, e^{-\frac{4}{18}i\pi}x, x_4,\ldots, x_{2n}\right) = 0\,.
    \end{align}
    These relations imply the following root structures
    \begin{align} \label{Eq:AppL2roots1}
        L_{2n}\left(e^{\frac{1}{18}i\pi}x, e^{-\frac{1}{18}i\pi}x, x_3,\ldots, x_{2n}\right) &\sim \prod_{k=3}^{2n}[11]_k[-11]_k\\  \label{Eq:AppL2roots2}
        L_{2n}\left(e^{\frac{5}{18}i\pi}x, e^{-\frac{5}{18}i\pi}x, x_3,\ldots, x_{2n}\right) &\sim \prod_{k=3}^{2n}[7]_k[-7]_k\\ \label{Eq:AppL2roots3}
        L_{2n}\left(e^{\frac{6}{18}i\pi}x, e^{-\frac{6}{18}i\pi}x, x_3,\ldots, x_{2n}\right) &\sim \prod_{k=3}^{2n}[4]_k[-4]_k\,.
    \end{align}
    We first remark that if
    \begin{align}
        L_{2n}\left(\varphi x, \varphi^{-1} x, x_3,\ldots, x_{2n}\right) = 0
    \end{align}
    for some symmetric homogeneous polynomial $L_{2n}$ and phase $\varphi$, then
    \begin{align}
        L_{2n}(x_1,\ldots, x_n) \sim \prod_{k<l}^{2n} (x_k - \varphi^2 x_l)(x_l - \varphi^2 x_k)\,,
    \end{align}
    which (combined with the kinematic part) exceeds the partial degree restrictions~\eqref{Eq:partDegSl} for any $n$. So we conclude that the polynomials eqs.~\eqref{Eq:AppL2roots1}, \eqref{Eq:AppL2roots2}, and \eqref{Eq:AppL2roots3} cannot be identically zero.
    
    Now we proceed to show that there is no such $L_{2n}$ that satisfies eqs.~\eqref{Eq:AppL2roots1}, \eqref{Eq:AppL2roots2}, and \eqref{Eq:AppL2roots3} and respects the partial degree restrictions, specifically
    \begin{align} \label{Eq:AppL2DegHigh}
        \deg_2\left(L_{2n}\right) \leq 4n - 3\,.
    \end{align}
    Eqs.~\eqref{Eq:AppL2roots1}, \eqref{Eq:AppL2roots2}, and \eqref{Eq:AppL2roots3} imply that the second partial degree of $L_{2n}$ is already at least $4n - 4$, so it is sufficient to show that if is necessary to add at least two further roots to satisfy all the three equations simultaneously. Consider $L_{2n}(x,y,x_3,\ldots,x_{2n})$ as a generic symmetric polynomial in $x$ and $y$:
    \begin{align}
        L_{2n}(x,y,x_3,\ldots,x_{2n}) = \sum_{k,l} C_{kl}\sigma_1^k\sigma_2^l\,,
    \end{align}
    where $\sigma_1 = x + y$, $\sigma_2 = xy$, and the $C_{kl}$ coefficients depend on $x_3,\ldots, x_{2n}$, and the sum is finite due to the partial degrees restrictions. Then
    \begin{align} \label{Eq:AppL2xPolyPhase}
        L_{2n}(e^{i \alpha} x,e^{-i \alpha} x,x_3,\ldots,x_{2n}) = \sum_{k,l} C_{kl} (2\cos \alpha)^k x^{k+2l}
    \end{align}
    is some polynomial of $x$. Its roots' sum and product can obtained in terms of its coefficients. Focusing on their $\alpha$-dependence we can write
    \begin{align} \label{Eq:AppSumRoots}
        \left(\sum \text{roots}\right) &= N (\cos \alpha)^q\quad\text{and}\\ \label{Eq:AppProdRoots}
        \left(\prod \text{roots}\right) &= \Tilde{N} (\cos \alpha)^{\Tilde{q}}
    \end{align}
    for some $q, \Tilde{q} \in \mathbb{Z}$.

    There are two sub-cases to consider.

    \subparagraph{Sub-case 3/A} First, assume that the polynomial in eq.~\eqref{Eq:AppL2xPolyPhase} has exactly the minimum number of roots, i.e., $4n - 4$ as prescribed by eqs.~\eqref{Eq:AppL2roots1}, \eqref{Eq:AppL2roots2}, and \eqref{Eq:AppL2roots3}. Then
    \begin{align}
        \left(\sum \text{roots}\right) &= 2\cos\phi_i\sum_{k=3}^{2n}x_k
    \end{align}
    where
    \begin{align}
        \phi_1 = \frac{11\pi}{18}\quad&\text{and}\quad\alpha_1 = \frac{\pi}{18}\,,\\
        \phi_2 = \frac{7\pi}{18}\quad&\text{and}\quad\alpha_2 = \frac{5\pi}{18}\,,\\
        \phi_3 = \frac{4\pi}{18}\quad&\text{and}\quad\alpha_3 = \frac{6\pi}{18}
    \end{align}
    corresponding to eqs.~\eqref{Eq:AppL2roots1}, \eqref{Eq:AppL2roots2}, and \eqref{Eq:AppL2roots3}, respectively. Then, we can take the ratio of two of the relations~\eqref{Eq:AppSumRoots} ($N \neq 0$ for the generic $x_3,\ldots, x_{2n}$) to get
    \begin{align}
        \frac{\cos\phi_i}{\cos\phi_j} = \left(\frac{\cos\alpha_i}{\cos\alpha_j}\right)^q
    \end{align}
    for every $i < j,\, \{i,j\}\subset \{1,2,3\}$. Now it is straightforward to check that this system of equations has no solution for $q$, therefore $4n - 4$ roots are not sufficient to satisfy eqs.~\eqref{Eq:AppL2roots1}, \eqref{Eq:AppL2roots2}, and \eqref{Eq:AppL2roots3} simultaneously.

    \subparagraph{Sub-case 3/B} Next, assume that the polynomial in eq.~\eqref{Eq:AppL2xPolyPhase} has exactly $4n - 3$ roots, that is, we allow one additional root, $r$ compared to the previous case. Then, the sum of the roots is
    \begin{align}
        \left(\sum \text{roots}\right) &= 2\cos\phi_i\sum_{k=3}^{2n}x_k + r\,.
    \end{align}
    There are two possibilities.
    \begin{itemize}
        \item First, we can choose
    \begin{align}
        r = -2\cos\phi_i\sum_{k=3}^{2n}x_k\,,
    \end{align}
    so the sum of the roots is zero. Consider the product of the roots
    \begin{align}
        \left(\prod \text{roots}\right) &= -2\cos\phi_i\left(\sum_{k=3}^{2n}x_k\right)\prod_{k=3}^{2n} x_k^2\,,
    \end{align}
    and take the ratio of two of the relations~\eqref{Eq:AppProdRoots} ($\Tilde{N} \neq 0$ for the generic $x_3,\ldots, x_{2n}$) to get
    \begin{align}
        \frac{\cos\phi_i}{\cos\phi_j} = \left(\frac{\cos\alpha_i}{\cos\alpha_j}\right)^{\Tilde{q}}
    \end{align}
    for every $i < j,\, \{i,j\}\subset \{1,2,3\}$. Similarly to the previous case, no $\Tilde{q}$ solves these equations simultaneously.
        \item Finally, we choose $r$ arbitrarily but assume that the sum of the roots does not vanish. From eq.~\eqref{Eq:AppSumRoots} we obtain
    \begin{align}
        \frac{\cos\phi_i + r'}{\cos\phi_j + r'} = \left(\frac{\cos\alpha_i}{\cos\alpha_j}\right)^q
    \end{align}
    for every $i < j,\, \{i,j\}\subset \{1,2,3\}$, where
    \begin{align}
        r' = \frac{r}{2\sum\limits_{k=3}^{2n}x_k}\,.
    \end{align}
    The system has two solutions in terms of $r'$ and $q$:
    \begin{align}
        q_1 &= 0\,,\quad r_1' = \infty\,,\quad \text{and}\\
        q_2 &\approx -0.15206\,,\quad r_2' \approx 10.5479\,.
    \end{align}
    On the one hand, since 
    \[
    \sum_{k=3}^{2n}x_k \neq 0
    \] 
    for the generic set of $x_k$'s, $r'$ can not be infinite, so the first solution is ruled out. On the other hand, the second solution is also not allowed, as $q$ must be an integer.
    
    \end{itemize}
    
    \noindent Therefore, to satisfy eqs.~\eqref{Eq:AppL2roots1}, \eqref{Eq:AppL2roots2}, and \eqref{Eq:AppL2roots3}, $L_{2n}(e^{i \alpha} x,e^{-i \alpha} x,x_3,\ldots,x_{2n})$ should have at least two additional roots, $4n - 4 + 2$ in total, and hence
    \begin{align}
        \deg_2\left(L_{2n}\right) \geq 4n - 2\,,
    \end{align}
    which is incompatible with eq.~\eqref{Eq:AppL2DegHigh}. Consequently, there is no symmetric homogeneous (non-zero) $L_{2n}$ polynomial that satisfies eqs.~\eqref{Eq:AppLZeroLHS} and \eqref{Eq:AppLZeroRHS} while respecting the partial degree restrictions.
    
    From Case 3, we learn that if a kernel of the kinematic singularity equation satisfies the $A_1^3\rightarrow A_1^3$ fusion and respects the partial degree restrictions, then Case 1 or 2 must apply. However, they only allow a nontrivial kernel solution for $2n \leq 11/2$ or $2n \leq 6$, respectively, which implies that there is no such kernel for $2n > 6$. Hence, the theorem is proven. 

    Note that the partial degree restriction of the $\Theta$ field yields even stronger constraints than the generic ones in~\eqref{Eq:partDegSl}.
\end{proof}

The lower bound estimation in the above proof is not the sharpest possible. One can check manually that there is no nontrivial kernel even for $2n = 6$. The marginal case is the four-particle form factor, where the kernel is precisely one dimensional, corresponding to the single free parameter that had to be fixed by the clustering property.

We further remark that exactly the same derivation works for the odd elements of the $A_1$ tower, guaranteeing the absence of a nontrivial kernel for $2n + 1 \geq 7$, while it can be shown manually that there is no kernel for $2n + 1 = 5$ either.

\clearpage

\bibliographystyle{utphys}
\bibliography{E7FF}

\providecommand{\href}[2]{#2}\begingroup\raggedright\begin{thebibliography}{10}

\bibitem{2010Sci...327..177C}
R.~{Coldea}, D.~A. {Tennant}, E.~M. {Wheeler}, E.~{Wawrzynska},
  D.~{Prabhakaran}, M.~{Telling}, K.~{Habicht}, P.~{Smeibidl}, and K.~{Kiefer},
  ``{Quantum Criticality in an Ising Chain: Experimental Evidence for Emergent
  E$_{8}$ Symmetry},'' \href{http://dx.doi.org/10.1126/science.1180085}{{\em
  Science} {\bfseries 327} (2010) 177},
  \href{http://arxiv.org/abs/1103.3694}{{\ttfamily arXiv:1103.3694
  [cond-mat.str-el]}}.

\bibitem{1989IJMPA...4.4235Z}
A.~B. {Zamolodchikov}, ``{Integrals of Motion and S-Matrix of the (scaled) $T =
  T_{c}$ Ising Model with Magnetic Field},''
  \href{http://dx.doi.org/10.1142/S0217751X8900176X}{{\em Int. J. Mod. Phys. A}
  {\bfseries 4} (1989) 4235--4248}.

\bibitem{2020PhRvB.102j4431A}
K.~{Amelin}, J.~{Engelmayer}, J.~{Viirok}, U.~{Nagel}, T.~{R{\~o}{\~o}m},
  T.~{Lorenz}, and Z.~{Wang}, ``{Experimental observation of quantum many-body
  excitations of $E_{8}$ symmetry in the Ising chain ferromagnet
  $\text{CoNb}_{2}\text{O}_{6}$},''
  \href{http://dx.doi.org/10.1103/PhysRevB.102.104431}{{\em Phys. Rev. B}
  {\bfseries 102} (2020) 104431},
  \href{http://arxiv.org/abs/2006.12956}{{\ttfamily arXiv:2006.12956
  [cond-mat.str-el]}}.

\bibitem{Amelin:2022ehz}
K.~Amelin, J.~Viirok, U.~Nagel, T.~R\~o\ om, J.~Engelmayer, T.~Dey,
  A.~Agung~Nugroho, T.~Lorenz, and Z.~Wang, ``{Quantum spin dynamics of
  quasi-one-dimensional Heisenberg-Ising magnets in a transverse field:
  confined spinons, E $_{8}$ spectrum, and quantum phase transitions},''
  \href{http://dx.doi.org/10.1088/1751-8121/aca6b8}{{\em J. Phys. A} {\bfseries
  55} no.~48, (2022) 484005}.

\bibitem{2024arXiv240211229G}
Y.~{Gao}, X.~{Wang}, N.~{Xi}, Y.~{Jiang}, R.~{Yu}, and J.~{Wu}, ``{Spin
  dynamics and dark particle in a weak-coupled quantum Ising ladder with
  $\mathcal{D}_8^{(1)}$ spectrum},''
  \href{http://arxiv.org/abs/2402.11229}{{\ttfamily arXiv:2402.11229
  [cond-mat.str-el]}}.

\bibitem{2024arXiv240310785X}
N.~{Xi}, X.~{Wang}, Y.~{Gao}, Y.~{Jiang}, R.~{Yu}, and J.~{Wu}, ``{Emergent
  $D_8^{(1)}$ spectrum and topological soliton excitation in CoNb$_2$O$_6$},''
  \href{http://arxiv.org/abs/2403.10785}{{\ttfamily arXiv:2403.10785
  [cond-mat.str-el]}}.

\bibitem{2020arXiv200513302Z}
H.~{Zou}, Y.~{Cui}, X.~{Wang}, Z.~{Zhang}, J.~{Yang}, G.~{Xu}, A.~{Okutani},
  M.~{Hagiwara}, M.~{Matsuda}, G.~{Wang}, G.~{Mussardo}, K.~{H{\'o}ds{\'a}gi},
  M.~{Kormos}, Z.~Z. {He}, S.~{Kimura}, R.~{Yu}, W.~{Yu}, J.~{Ma}, and J.~{Wu},
  ``{$E_8$ Spectra of Quasi-one-dimensional Antiferromagnet BaCo$_2$V$_2$O$_8$
  under Transverse Field},''
  \href{http://dx.doi.org/10.1103/PhysRevLett.127.077201}{{\em Phys. Rev.
  Lett.} {\bfseries 127} (2021) 077201},
  \href{http://arxiv.org/abs/2005.13302}{{\ttfamily arXiv:2005.13302
  [cond-mat.str-el]}}.

\bibitem{2021PhRvB.103w5117W}
X.~{Wang}, H.~{Zou}, K.~{H{\'o}ds{\'a}gi}, M.~{Kormos}, G.~{Tak{\'a}cs}, and
  J.~{Wu}, ``{Cascade of singularities in the spin dynamics of a perturbed
  quantum critical Ising chain},''
  \href{http://dx.doi.org/10.1103/PhysRevB.103.235117}{{\em Phys. Rev. B}
  {\bfseries 103} (2021) 235117},
  \href{http://arxiv.org/abs/2103.09128}{{\ttfamily arXiv:2103.09128
  [cond-mat.str-el]}}.

\bibitem{2023arXiv230800249W}
X.~{Wang}, K.~{Puzniak}, K.~{Schmalzl}, C.~{Balz}, M.~{Matsuda}, A.~{Okutani},
  M.~{Hagiwara}, J.~{Ma}, J.~{Wu}, and B.~{Lake}, ``{Spin dynamics of the $E_8$
  particles},'' \href{http://arxiv.org/abs/2308.00249}{{\ttfamily
  arXiv:2308.00249 [cond-mat.str-el]}}.

\bibitem{2023JPhA...56L3001L}
M.~{Lencs{\'e}s}, G.~{Mussardo}, and G.~{Tak{\'a}cs}, ``{Quantum integrability
  vs experiments: correlation functions and dynamical structure factors},''
  \href{http://dx.doi.org/10.1088/1751-8121/acf255}{{\em J. Phys: A Math. Gen.}
  {\bfseries 56} (2023) 383001},
  \href{http://arxiv.org/abs/2303.16556}{{\ttfamily arXiv:2303.16556
  [cond-mat.stat-mech]}}.

\bibitem{Smirnov}
F.~A. Smirnov, \href{http://dx.doi.org/10.1142/1115}{{\em {Form Factors in
  Completely Integrable Models of Quantum Field Theory}}}.
\newblock World Scientific, 1992.

\bibitem{Delfino:1996jr}
G.~Delfino and P.~Simonetti, ``{Correlation functions in the two-dimensional
  Ising model in a magnetic field at $T = T_c$},''
  \href{http://dx.doi.org/10.1016/0370-2693(96)00783-6}{{\em Phys. Lett. B}
  {\bfseries 383} (1996) 450--456},
  \href{http://arxiv.org/abs/hep-th/9605065}{{\ttfamily arXiv:hep-th/9605065}}.

\bibitem{DMIMMF}
G.~{Delfino} and G.~{Mussardo}, ``{The spin-spin correlation function in the
  two-dimensional Ising model in a magnetic field at $T = T_{c}$},''
  \href{http://dx.doi.org/10.1016/0550-3213(95)00464-4}{{\em Nucl. Phys. B}
  {\bfseries 455} (1995) 724--758},
  \href{http://arxiv.org/abs/hep-th/9507010}{{\ttfamily arXiv:hep-th/9507010
  [hep-th]}}.

\bibitem{MC}
P.~{Christe} and G.~{Mussardo}, ``{Integrable systems away from critically: The
  Toda field theory and S-matrix of the tricritical Ising model},''
  \href{http://dx.doi.org/10.1016/0550-3213(90)90119-X}{{\em Nucl. Phys. B}
  {\bfseries 330} (1990) 465--487}.

\bibitem{FZ}
V.~A. Fateev and A.~B. Zamolodchikov, ``{Conformal field theory and purely
  elastic S matrices},''
  \href{http://dx.doi.org/10.1142/S0217751X90000477}{{\em Int. J. Mod. Phys. A}
  {\bfseries 5} (1990) 1025--1048}.

\bibitem{AMV}
C.~{Acerbi}, A.~{Valleriani}, and G.~{Mussardo}, ``{Form Factors and
  Correlation Functions of the Stress-Energy Tensor in Massive Deformation of
  the Minimal Models $(E_{n})_{1} \otimes (E_{n})_{1}/(E_{n})_{2}$},''
  \href{http://dx.doi.org/10.1142/S0217751X96002443}{{\em Int. J. Mod. Phys. A}
  {\bfseries 11} (1996) 5327--5364},
  \href{http://arxiv.org/abs/hep-th/9601113}{{\ttfamily arXiv:hep-th/9601113
  [hep-th]}}.

\bibitem{2022ScPP...12..162C}
A.~{Cortes Cubero}, R.~{Konik}, M.~{Lencs{\'e}s}, G.~{Mussardo}, and
  G.~{Tak{\'a}cs}, ``{Duality and form factors in the thermally deformed
  two-dimensional tricritical Ising model},''
  \href{http://dx.doi.org/10.21468/SciPostPhys.12.5.162}{{\em SciPost Phys.}
  {\bfseries 12} (2022) 162}, \href{http://arxiv.org/abs/2109.09767}{{\ttfamily
  arXiv:2109.09767 [cond-mat.stat-mech]}}.

\bibitem{2024JSMTE2024c3103M}
G.~{Mussardo}, M.~{Panero}, and A.~{Stampiggi}, ``{Form factors of the
  tricritical three-state Potts model in its scaling limit},''
  \href{http://dx.doi.org/10.1088/1742-5468/ad2926}{{\em J. Stat. Mech. Theor.
  Exp.} {\bfseries 2024} (2024) 033103},
  \href{http://arxiv.org/abs/2311.00654}{{\ttfamily arXiv:2311.00654
  [hep-th]}}.

\bibitem{2019PhRvL.123i0401B}
A.~S. {Buyskikh}, L.~{Tagliacozzo}, D.~{Schuricht}, C.~A. {Hooley},
  D.~{Pekker}, and A.~J. {Daley}, ``{Spin Models, Dynamics, and Criticality
  with Atoms in Tilted Optical Superlattices},''
  \href{http://dx.doi.org/10.1103/PhysRevLett.123.090401}{{\em Phys. Rev.
  Lett.} {\bfseries 123} (2019) 090401},
  \href{http://arxiv.org/abs/1811.06995}{{\ttfamily arXiv:1811.06995
  [cond-mat.quant-gas]}}.

\bibitem{2021PhRvB.104w5109S}
K.~{Slagle}, D.~{Aasen}, H.~{Pichler}, R.~S.~K. {Mong}, P.~{Fendley},
  X.~{Chen}, M.~{Endres}, and J.~{Alicea}, ``{Microscopic characterization of
  Ising conformal field theory in Rydberg chains},''
  \href{http://dx.doi.org/10.1103/PhysRevB.104.235109}{{\em Phys. Rev. B}
  {\bfseries 104} (2021) 235109},
  \href{http://arxiv.org/abs/2108.09309}{{\ttfamily arXiv:2108.09309
  [cond-mat.str-el]}}.

\bibitem{2022PhRvB.106k5122S}
K.~{Slagle}, Y.~{Liu}, D.~{Aasen}, H.~{Pichler}, R.~S.~K. {Mong}, X.~{Chen},
  M.~{Endres}, and J.~{Alicea}, ``{Quantum spin liquids bootstrapped from Ising
  criticality in Rydberg arrays},''
  \href{http://dx.doi.org/10.1103/PhysRevB.106.115122}{{\em Phys. Rev. B}
  {\bfseries 106} (2022) 115122},
  \href{http://arxiv.org/abs/2204.00013}{{\ttfamily arXiv:2204.00013
  [cond-mat.str-el]}}.

\bibitem{2023PhRvB.108w5414R}
A.~{Roy}, ``{Quantum electronic circuits for multicritical Ising models},''
  \href{http://dx.doi.org/10.1103/PhysRevB.108.235414}{{\em Phys. Rev. B}
  {\bfseries 108} (2023) 235414},
  \href{http://arxiv.org/abs/2306.04346}{{\ttfamily arXiv:2306.04346
  [quant-ph]}}.

\bibitem{PhysRevLett.132.226502}
L.~Maffi, N.~Tausendpfund, M.~Rizzi, and M.~Burrello, ``Quantum simulation of
  the tricritical ising model in tunable josephson junction ladders,''
  \href{http://dx.doi.org/10.1103/PhysRevLett.132.226502}{{\em Phys. Rev.
  Lett.} {\bfseries 132} (2024) 226502},
  \href{http://arxiv.org/abs/2310.18300}{{\ttfamily arXiv:2310.18300}}.

\bibitem{2022PhLB..82837008L}
M.~{Lencs{\'e}s}, G.~{Mussardo}, and G.~{Tak{\'a}cs}, ``{Confinement in the
  tricritical Ising model},''
  \href{http://dx.doi.org/10.1016/j.physletb.2022.137008}{{\em Phys. Lett. B}
  {\bfseries 828} (2022) 137008},
  \href{http://arxiv.org/abs/2111.05360}{{\ttfamily arXiv:2111.05360
  [hep-th]}}.

\bibitem{2022PhRvD.106j5003L}
M.~{Lencs{\'e}s}, G.~{Mussardo}, and G.~{Tak{\'a}cs}, ``{Variations on vacuum
  decay: The scaling Ising and tricritical Ising field theories},''
  \href{http://dx.doi.org/10.1103/PhysRevD.106.105003}{{\em Phys. Rev. D}
  {\bfseries 106} (2022) 105003},
  \href{http://arxiv.org/abs/2208.02273}{{\ttfamily arXiv:2208.02273
  [hep-th]}}.

\bibitem{BC1}
M.~Blume, ``{Theory of the First-Order Magnetic Phase Change in
  U${\mathrm{O}}_{2}$},'' \href{http://dx.doi.org/10.1103/PhysRev.141.517}{{\em
  Phys. Rev.} {\bfseries 141} (1966) 517--524}.

\bibitem{BC2}
H.~W. {Capel}, ``{On the possibility of first-order phase transitions in Ising
  systems of triplet ions with zero-field splitting},''
  \href{http://dx.doi.org/10.1016/0031-8914(66)90027-9}{{\em Physica}
  {\bfseries 32} (1966) 966--988}.

\bibitem{vonGehlen:1989yn}
G.~von Gehlen, ``{Off Criticality Behavior of the Blume-Capel Quantum Chain as
  a Check of Zamolodchikov's Conjecture},''
  \href{http://dx.doi.org/10.1016/0550-3213(90)90130-6}{{\em Nucl. Phys. B}
  {\bfseries 330} (1990) 741--756}.

\bibitem{1979AnPhy.120..253Z}
A.~B. {Zamolodchikov} and A.~B. {Zamolodchikov}, ``{Factorized S-matrices in
  two dimensions as the exact solutions of certain relativistic quantum field
  theory models},'' \href{http://dx.doi.org/10.1016/0003-4916(79)90391-9}{{\em
  Annals Phys.} {\bfseries 120} (1979) 253--291}.

\bibitem{FATEEV199445}
V.~Fateev, ``{The exact relations between the coupling constants and the masses
  of particles for the integrable perturbed conformal field theories},''
  \href{http://dx.doi.org/10.1016/0370-2693(94)00078-6}{{\em Phys. Lett. B}
  {\bfseries 324} (1994) 45--51}.

\bibitem{Koubek:1993ke}
A.~Koubek and G.~Mussardo, ``{On the operator content of the sinh-Gordon
  model},'' \href{http://dx.doi.org/10.1016/0370-2693(93)90554-U}{{\em Phys.
  Lett. B} {\bfseries 311} (1993) 193--201},
  \href{http://arxiv.org/abs/hep-th/9306044}{{\ttfamily arXiv:hep-th/9306044}}.

\bibitem{AMVcluster}
C.~{Acerbi}, G.~{Mussardo}, and A.~{Valleriani}, ``{On the form factors of
  relevant operators and their cluster property},''
  \href{http://dx.doi.org/10.1088/0305-4470/30/9/007}{{\em J. Phys. A Math.
  Gen.} {\bfseries 30} (1997) 2895--2913},
  \href{http://arxiv.org/abs/hep-th/9609080}{{\ttfamily arXiv:hep-th/9609080
  [hep-th]}}.

\bibitem{DSC}
G.~{Delfino}, P.~{Simonetti}, and J.~L. {Cardy}, ``{Asymptotic factorisation of
  form factors in two-dimensional quantum field theory},''
  \href{http://dx.doi.org/10.1016/0370-2693(96)01035-0}{{\em Phys. Lett. B}
  {\bfseries 387} (1996) 327--333},
  \href{http://arxiv.org/abs/hep-th/9607046}{{\ttfamily arXiv:hep-th/9607046
  [hep-th]}}.

\bibitem{1997NuPhB.497..589A}
C.~{Acerbi}, ``{Form factors of exponential operators and exact wave function
  renormalization constant in the Bullough-Dodd model},''
  \href{http://dx.doi.org/10.1016/S0550-3213(97)00303-9}{{\em Nucl. Phys. B}
  {\bfseries 497} (1997) 589--610},
  \href{http://arxiv.org/abs/hep-th/9701062}{{\ttfamily arXiv:hep-th/9701062
  [hep-th]}}.

\bibitem{PhysRevD.19.2477}
B.~Berg, M.~Karowski, and P.~Weisz, ``{Construction of Green's functions from
  an exact $S$ matrix},''
  \href{http://dx.doi.org/10.1103/PhysRevD.19.2477}{{\em Phys. Rev. D}
  {\bfseries 19} (1979) 2477--2479}.

\bibitem{russian2}
V.~{Fateev}, S.~{Lukyanov}, A.~{Zamolodchikov}, and A.~{Zamolodchikov},
  ``{Expectation values of local fields in the Bullough-Dodd model and
  integrable perturbed conformal field theories},''
  \href{http://dx.doi.org/10.1016/S0550-3213(98)00002-9}{{\em Nucl. Phys. B}
  {\bfseries 516} (1998) 652--674},
  \href{http://arxiv.org/abs/hep-th/9709034}{{\ttfamily arXiv:hep-th/9709034
  [hep-th]}}.

\bibitem{1991NuPhB.348..619Z}
A.~B. {Zamolodchikov}, ``{Two-point correlation function in scaling Lee-Yang
  model},'' \href{http://dx.doi.org/10.1016/0550-3213(91)90207-E}{{\em Nucl.
  Phys. B} {\bfseries 348} (1991) 619--641}.

\bibitem{1993NuPhB.393..413F}
A.~{Fring}, G.~{Mussardo}, and P.~{Simonetti}, ``{Form factors for integrable
  lagrangian field theories, the sinh-Gordon model},''
  \href{http://dx.doi.org/10.1016/0550-3213(93)90252-K}{{\em Nucl. Phys. B}
  {\bfseries 393} (1993) 413--441},
  \href{http://arxiv.org/abs/hep-th/9211053}{{\ttfamily arXiv:hep-th/9211053
  [hep-th]}}.

\bibitem{1997hep.th....5142P}
M.~{Pillin}, ``{Polynomial Recursion Equations in Form Factors of ADE Toda
  Field Theories},''
  \href{http://dx.doi.org/10.48550/arXiv.hep-th/9705142}{{\em Lett. Math.
  Phys.} {\bfseries 43} (1998) 211--224},
  \href{http://arxiv.org/abs/hep-th/9705142}{{\ttfamily arXiv:hep-th/9705142
  [hep-th]}}.

\end{thebibliography}\endgroup

\end{document}